\documentclass[a4paper,fleqn,usenatbib]{mnras}

\usepackage{newtxtext,newtxmath}

\usepackage[T1]{fontenc}
\usepackage{ae,aecompl}

\usepackage{graphicx}	
\usepackage{amsmath}	
\usepackage{amssymb}	
\usepackage{booktabs, siunitx}    
\usepackage{enumerate}  
\usepackage{nicefrac}  

\newcommand\Tstrut{\rule{0pt}{2.6ex}}       
\newcommand\Bstrut{\rule[-1.1ex]{0pt}{0pt}}
\def\CIII{C~\textsc{iii}]}

\title[The jet of the $\gamma$-NLS1 PKS\,J1222$+$0413]{The relativistic jet of the $\gamma$-ray emitting narrow-line Seyfert 1 galaxy PKS\,J1222$+$0413}

\author[D. Kynoch et al.]{
Daniel Kynoch,$^{1}$\thanks{E-mail: daniel.kynoch@durham.ac.uk}
Hermine Landt$^{1}$,
Martin J. Ward$^{1}$,
Chris Done$^{1}$,
\newauthor
Catherine Boisson$^{2}$,
Mislav Balokovi\'{c}$^{3}$,
Emmanouil Angelakis$^{4}$ and
\newauthor
Ioannis Myserlis$^{4}$
\\
$^{1}$Centre for Extragalactic Astronomy, Department of Physics, Durham University, South Road, Durham, DH1 3LE, UK \\
$^{2}$LUTH, Observatoire de Paris, CNRS, Universit\'{e} Paris Diderot, PSL Research University Paris, 5 place Jules Janssen, \\92195 Meudon, France\\
$^{3}$Cahill Center for Astronomy and Astrophysics, California Institute of Technology, Pasadena, CA 91125, USA \\
$^{4}$Max-Planck-Institut f\"{u}r Radioastronomie, Auf dem H\"{u}gel 69, 53121 Bonn, Germany
}

\date{Accepted XXX. Received YYY; in original form ZZZ}

\pubyear{2019}

\begin{document}
\label{firstpage}
\pagerange{\pageref{firstpage}--\pageref{lastpage}}
\maketitle

\begin{abstract}
We present a multi-frequency study of PKS\,J1222$+$0413 (4C\,$+04.42$), currently the highest redshift $\gamma$-ray emitting narrow-line Seyfert 1 ($\gamma$-NLS1).
We assemble a broad spectral energy distribution (SED) including previously unpublished datasets: X-ray data obtained with the \textit{NuSTAR} and \textit{Neil Gehrels Swift} observatories; near-infrared, optical and UV spectroscopy obtained with VLT X-shooter; and multiband radio data from the Effelsberg telescope.  
These new observations are supplemented by archival data from the literature.
We apply physical models to the broadband SED, parameterising the accretion flow and jet emission to investigate the disc-jet connection.
PKS\,J1222$+$0413 has a much greater black hole mass than most other NLS1s, $M_\mathrm{BH}\approx{\color{black}2}\times10^8$~M$_{\sun}$, similar to those found in flat spectrum radio quasars (FSRQs).
Therefore this source provides insight into how the jets of $\gamma$-NLS1s relate to those of FSRQs. 
\end{abstract}




\section{Introduction}
The relativistic jets present in a subset of active galactic nuclei (AGN) are among the most energetic phenomena in the Universe.
Matter transported in these jets is expelled from the nucleus at velocities approaching the speed of light.
These jets are highly-collimated and can extend for great distances beyond the galaxy in which they originate, far out into intergalactic space.
The most spectacular examples of these are the jets of blazars, in which the jet is aligned to our line of sight.
As a result of relativistic beaming, the emission from the jet is amplified and non-thermal emission from the jet can be detected across the whole electromagnetic spectrum from radio to $\gamma$-rays.

Blazar jet emission is understood to originate from highly energetic electrons within the jet.
Synchrotron emission resulting from the electrons entrained in magnetic field lines in the jet produces a low-frequency peak (at radio-to-X-ray frequencies) in the spectral energy distribution (SED).
A second, higher-frequency, peak (at X-ray-to-$\gamma$-ray frequencies) results from the up-scattering of soft seed photons; the peak frequency and luminosity of this feature depends on the environment in which the electrons cool.
Flat spectrum radio quasars (FSRQs) are high accretion-rate systems ($L/L_\mathrm{Edd}\sim0.1$) with a luminous accretion disc and broad line region (BLR).
In FSRQs, electrons within the jet cool by up-scattering seed photons originating in the accretion flow (from the disc and its corona, the BLR and extended, dusty torus) via the external Compton (EC) mechanism (e.g.\ \citealt{Gardner18}).
In contrast, the dearth of external seed photons in low accretion rate BL Lacertae objects (BL Lacs) means the electrons cool by Compton upscattering the synchrotron photons they emit, via the synchrotron self-Compton (SSC) process (e.g.\ \citealt{Gardner14}, \citealt{G&T09}).
BL Lacs and FSRQs form a `blazar sequence' (\citealt{Fossati98}, \citealt{Ghisellini17}) from the low-power, high-frequency peaked SEDs of BL Lacs to high-power, low-frequency peaked SEDs of FSRQs. 

\cite{Ghisellini14} found that the jet power of FSRQs and BL Lacs is correlated with, and often exceeds, the accretion power.
It has been proposed that jets are able to draw power from the angular momentum of spinning black holes (BHs) (\citealt{BZ77}) or may be launched by a high concentration of accumulated magnetic flux (\citealt{Sikora13}), although the launching and powering mechanisms of jets are still debated.
Some early studies thought that the mass of the BH must play a role, since FSRQs and BL Lacs all have high-mass BHs ($\log(M_\mathrm{BH}/\mathrm{M}_{\sun})\gtrsim8$--9, \citealt{G&T10}) and are hosted in large, elliptical galaxies.
This view changed with the \textit{Fermi Gamma-Ray Space Telescope} detection of $\gamma$-ray emission from narrow-line Seyfert 1 galaxies (\citealt{Abdo09b}).
NLS1s usually have low BH masses $\langle\log(M_\mathrm{BH}/\mathrm{M}_{\sun})\rangle=6.9$, high accretion rates (\citealt{Rakshit17}) and are typically found in spiral galaxies.
With luminous accretion discs and BLRs, the electron cooling environment of their jets (and hence the spectral properties of their non-thermal SEDs) are very similar to those of FSRQs, but at much lower jet powers.
However, where (and whether) $\gamma$-NLS1s fit on the standard blazar sequence is still an open question.
By investigating the jet and accretion flow properties of $\gamma$-NLS1s we can better understand how relativistic jets scale with the mass and accretion rate of the central BH. 

\subsection{The source PKS\,J1222$+$0413}
PKS\,J1222$+$0413 (RA: 12~22~22.548, Dec: $+$04~13~15.75) was identified as a $\gamma$-ray NLS1 candidate by \cite{Yao15} and, at redshift $z=0.9662$, is currently the most distant example of this class known.
From its SDSS spectrum, they determined its mass to be $M_\mathrm{BH}=1.8$--$2.0\times10^{8}$~M$_{\sun}$.
Its simultaneous and quasi-simultaneous SED was studied by \cite{Giommi12}, although at the time it was not known to be a NLS1.
Like 1H\,0323$+$342, which was studied in great detail by \cite{Kynoch18}, the accretion disc of PKS\,J1222$+$0413 is a prominent feature in its spectral energy distribution, unlike other $\gamma$-NLS1s which are more jet-dominated. 
PKS\,J1222$+$0413 is therefore another ideal object in which to examine the disc-jet connection.
We present new data from \textit{NuSTAR}, VLT X-shooter and Effelsberg.
The VLT X-shooter data allows us to investigate the NLS1 nature of the source and measure the AGN continuum emission from the accretion disc.
The 0.3--10~keV X-ray band has been shown to contain contributions from both the accretion flow and jet (e.g.\ \citealt{Kynoch18}, \citealt{Larsson18}) whereas the higher-energy X-rays, sampled by \textit{NuSTAR} and \textit{Swift} BAT, are jet-dominated.

Throughout this paper, we assume a $\Lambda$CDM cosmology with $H_0=70$~km~s$^{-1}$~Mpc$^{-1}$, $\Omega_\mathrm{m}=0.3$ and $\Omega_\Lambda=0.7$.
Therefore the redshift $z=0.9662$ implies a luminosity distance of 6332.3~Mpc and a flux-to-luminosity conversion factor of $4.80\times10^{57}$~cm$^2$.
Length scales are often quoted in gravitational radii, $R_\mathrm{g}=GM_\mathrm{BH}/c^2={\color{black}2.95}\times10^{11}$~m $={\color{black}1.14}\times10^{-2}$~light days for our derived mass.

\section{X-Shooter spectroscopy}
\subsection{The observations and data reduction}
We have conducted X-shooter (\citealt{Xshooter}) observations of five $\gamma$-NLS1s (including PKS\,J1222$+$0413) in the Southern and equatorial sky, with the aim of characterising their accretion disc continuum and emission line properties.
Our near-infrared, optical and ultraviolet spectra were obtained using X-shooter which is a wide-band, intermediate resolution spectrograph developed for the Very Large Telescope (VLT).
The instrument is mounted on the VLT's 8.2~m Unit Telescope 2 at ESO Paranal in Chile.
Dichroics in the instrument reflect light into ultraviolet-blue (UVB) and visible light (VIS) arms and transmit the remaining light to the near-infrared (NIR) arm.
At the end of each arm is an \'{e}chelle spectrograph, with the UVB spectrograph covering the 300--550~nm range, the VIS spectrograph covering 550-1010~nm and the NIR spectrograph covering 1000--2500~nm.
Thus, by operating these three spectrographs simultaneously, X-shooter offers a very wide wavelength coverage from 300 to 2500~nm. 
We used the $0.9^{\prime\prime}\times11^{\prime\prime}$, $1.2^{\prime\prime}\times11^{\prime\prime}$, $1.3^{\prime\prime}\times11^{\prime\prime}$ slits, giving spectral resolutions $R\approx5600$, 6700 and 4000 in the IR, optical and UV spectra, respectively. 
The observation was conducted on 3 April 2017 under favourable weather conditions with clear skies and seeing $\approx1^{\prime\prime}$ at 500~nm.
The on-source integration time was 1 hour and the mean S/N achieved in the IR, optical and UV spectra were 12, 16 and 23, respectively. 

The data reduction was performed with the ESO \textsc{reflex} software which outputs wavelength- and flux-calibrated 1D spectra.
Corrections for telluric absorption in the optical and UV spectra were performed with the xtellcor\_general tool available as part of the \textsc{spextool} IDL package (\citealt{Spextool1}; \citealt{Spextool2}).

To sample the AGN continuum emission within the X-shooter spectral range, we averaged the flux in several emission line free windows of width 50~\AA.
These were (rest frame): 1660, 2000 and 2200~\AA\ observed in the UV; 3050, 3930 and 4205~\AA\ observed in the optical; 5100, 6205, 8600 and 11020~\AA\ observed in the IR. 
Five of these windows are those suggested by \cite{Capellupo15} for the purpose of modelling accretion disc emission in SEDs. 

\begin{figure*}
\centering
\includegraphics[width=1.5\columnwidth, keepaspectratio, angle=-90]{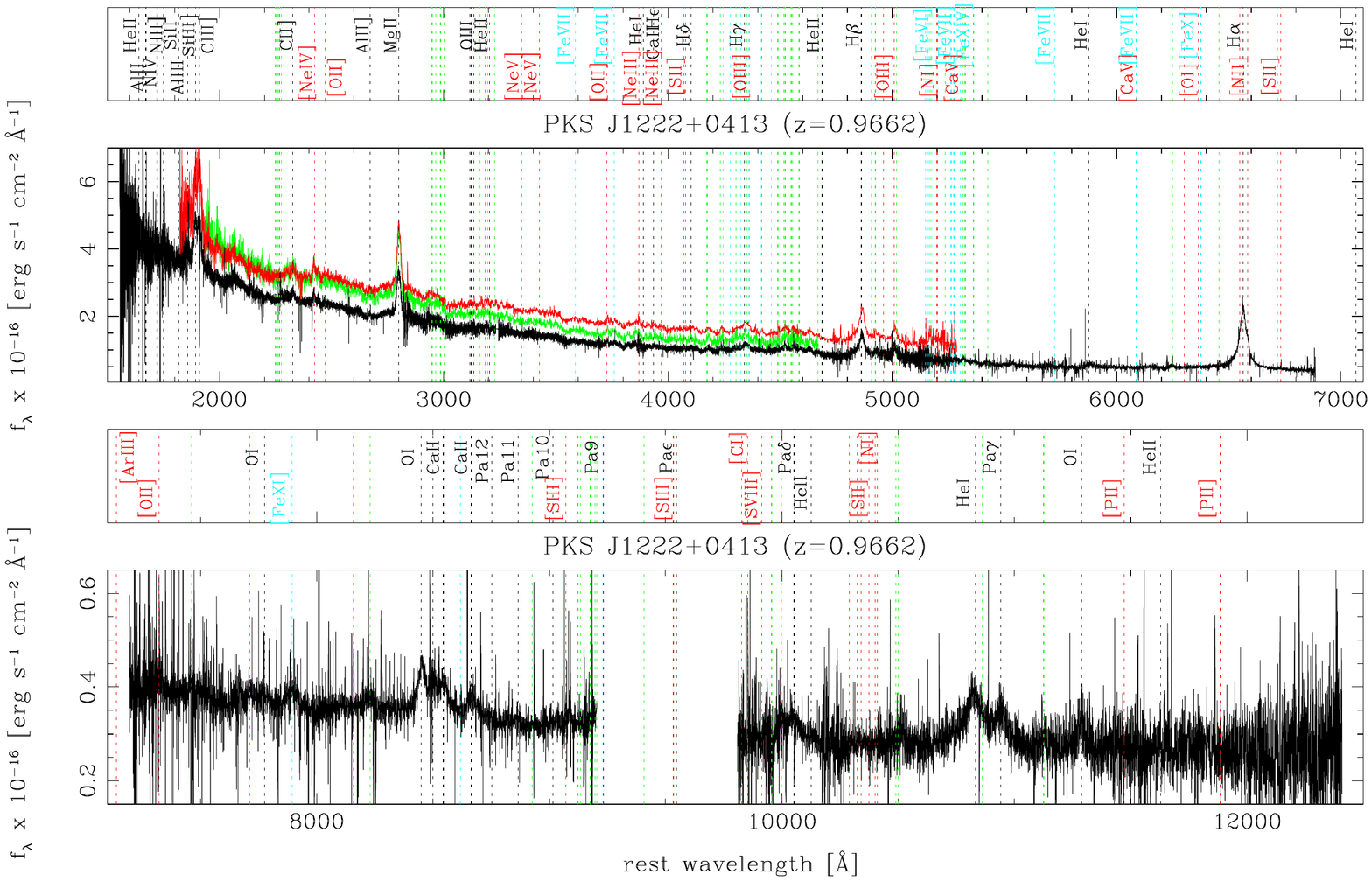}
\caption[General Caption]{VLT X-shooter spectra (black) taken 3 April 2017.  SDSS spectra taken 13 February 2008 and 26 March 2011 are overplotted in green and red, respectively.}
\label{fig:xsh}
\end{figure*}

\subsection{Estimates of the black hole mass}
\label{sec:bhmass}
We estimate the BH mass of PKS\,J1222$+$0413 
{\color{black} using} virial scaling relations employing the full widths at half-maximum (FWHMs) of the broad components of permitted emission lines (namely H$\upalpha$, H$\upbeta$ and Mg\,\textsc{ii}) and the ionising continuum luminosities measured at 5100 and 3000~\AA\ in the rest frame.
All of these quantities are covered simultaneously by our
X-shooter spectrum. The measurements of the 5100~\AA~and
3000~\AA~continuum luminosities are straightforward and we get values
of $\log \nu L_{\rm 5100\AA}=45.56$~erg~s$^{-1}$ and $\log \nu L_{\rm
3000\AA}=45.66$~erg~s$^{-1}$, respectively. 

\begin{figure*}
\centerline{
\includegraphics[width=1.5\columnwidth, keepaspectratio]{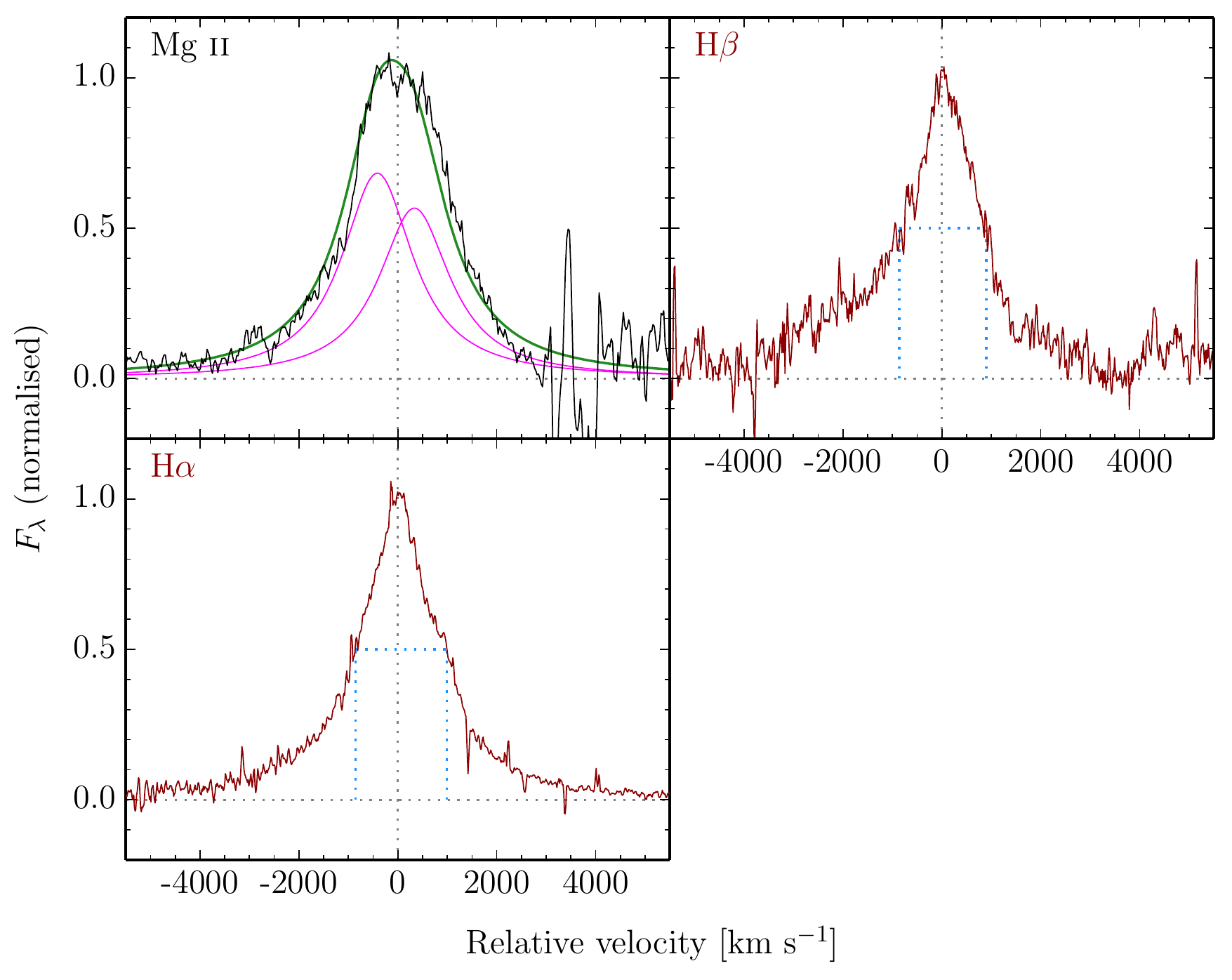}
}
\caption{\label{Hgam} 
The profiles of the three broad emission lines H$\upalpha$, H$\upbeta$
and Mg\,\textsc{ii}~that we used to estimate the BH mass for
PKS~J1222$+$0413. The profiles are shown in velocity space relative to
the expected rest-frame wavelength and have been Fe\,\textsc{ii}- and continuum-subtracted
and normalized to the same peak intensity.
In the top left panel, the Mg\,\textsc{ii} profile has been modelled as the sum of two 
Lorentzians (magenta) and the total modelled profile is shown in green.
Since the narrow line region likely makes a negligible contribution to the
Balmer line fluxes, the FWHM of the broad H$\upbeta$ and H$\upalpha$ 
lines may be estimated directly from the observed line profiles, as shown.
}
\end{figure*}

Before measuring the emission line widths and fluxes we first
modelled and subtracted Fe\,\textsc{ii} emission complexes around
H$\upbeta$ and Mg\,\textsc{ii} following the method described in \cite{Landt17}.
{\color{black}
The Mg\,\textsc{ii}$\lambda2008$ emission line is a blended doublet of lines centred at 2795 and 2802~\AA\ with an intensity
ratio of approximately 1.2:1.
To determine the widths of the individual broad lines we model the doublet as the sum of 
two broad Lorentzians with their intensities and wavelength separation fixed
to the theoretical values.  Additionally, we include two narrow Gaussian lines
to account for any contribution from the narrow emission line region.  The narrow lines
are fixed to the same intensity ratio and separation, and their widths are limited to be in 
the range $300\leqslant\mathrm{FWHM}\leqslant900$~km~s$^{-1}$.
The fitting is performed with a custom \textsc{python} script employing the
\textsc{lmfit} package, which uses the Levenberg-Marquardt method of least-squares
fitting.  We find that the inclusion of the narrow lines does not improve the fit
therefore we model the doublet simply as the sum of two Lorentzians (see Figure~\ref{Hgam}).  
The FWHM of these lines is $\approx1750$~km~s$^{-1}$.
}
Using the recently-recalibrated 
BH mass relationship of \citet{Mejia16}\footnote{Specifically, we use the calibration 
for the local continuum fit corrected for small systematic offsets (see their Table 7).}, 
we estimate the BH mass to be $\left({\color{black}2.6^{+2.0}_{-1.2}}\right)\times10^8$~M$_{\sun}$. 

In order to correctly measure the widths of the Balmer emission line
broad components, we need to subtract any contribution from the narrow line region.
{\color{black}
Commonly, the [O\,\textsc{iii}]$\lambda5007$ forbidden emission line is 
taken as being representative of the narrow emission line profiles.  A narrow line
with the same profile as [O\,\textsc{iii}]$\lambda5007$ 
is then subtracted from the (total) permitted line profiles, in principle leaving only
their broad components.
However, the [O\,\textsc{iii}]$\lambda5007$ line is very weak 
so its profile is unsuitable for this purpose.
Instead, we measure its total flux and use this to estimate the likely scaled flux of
narrow H$\upbeta$ in the following way.
`Normal' Seyferts have a typical [O\,\textsc{iii}]$\lambda5007$ to narrow 
H$\upbeta$ flux ratio of $\approx10$ (e.g.\ \citealt{Cohen83}) and
analyses of large samples have found consistent narrow line ratios in NLS1s 
(e.g.\ \citealt{Veron-Cetty01}; \citealt{Zhou06}; \citealt{Rakshit17}).
We estimate the [O\,\textsc{iii}]$\lambda5007$ line luminosity by integrating the observed flux density
 and we determine $\log L_{\mathrm{[O\textsc{iii}]}}=42.3$~erg~s$^{-1}$.
Assuming the ratio $L_{\mathrm{[O\textsc{iii}]}}/L^{\mathrm{n}}_{\mathrm{H}\upbeta}=10$ 
for our source, $\log L^{\mathrm{n}}_{\mathrm{H\upbeta}}=41.3$~erg~s$^{-1}$, which 
is negligible compared to the total line luminosity $\log L_{\rm H\upbeta}=43.11$~erg~s$^{-1}$.
Clearly, if the contribution from the narrow line component is negligible then we can measure the FWHM of H$\upbeta$ 
directly from its observed profile and we obtain $\mathrm{FWHM(H\upbeta)}\approx1760$~km~s$^{-1}$, which is 
extremely close to the FWHMs of the blended components of the Mg\,\textsc{ii} doublet. 
}
Using 
the \cite{Bentz13} radius-luminosity relationship for the H$\upbeta$
line\footnote{\color{black}Specifically, their `Clean2$+$ExtCorr' calibration.}, 
derived from optical reverberation mapping results, we determine
a BLR radius of $260^{+91}_{-67}$~light-days. Then, assuming a geometrical scaling 
factor of $f=1.4$ appropriate for FWHM measures \citep{Onken04}, we derive a BH mass of
$\left({\color{black}2.2^{+0.8}_{-0.5}}\right)\times10^8$~M$_{\sun}$.
We measure a $\mathrm{FWHM}={\color{black}1850}$~km~s$^{-1}$ for H$\upalpha$. Using 
the BH mass relationship of \citet{Mejia16} we
estimate the BH mass to be $\left({\color{black}2.5^{+1.1}_{-0.8}}\right)\times10^8$~M$_{\sun}$.   
These mass estimates and the relations used to calculate them are tabulated in Table 
\ref{tab:bhmass}.


\begin{table*}
\caption{\label{tab:bhmass} Estimates of the black hole mass}
\begin{tabular}{lcl}
\hline
Measurements & $M_{\rm BH}$        & Reference \\
             & (10$^8$ M$_{\sun}$)  & \\
\hline
$\log \nu L_{\rm 3000\AA}=45.66$~erg~s$^{-1}$ 					& {\color{black}2.6$^{+2.0}_{-1.2}$} & Table 7 of \citet{Mejia16} \\
FWHM(Mg\,\textsc{ii})={\color{black}1750}~km~s$^{-1}$             & & \\ \Tstrut\Bstrut
$\log \nu L_{\rm 5100\AA}=45.56$~erg~s$^{-1}$ 					& {\color{black}2.5$^{+1.1}_{-0.8}$} & Table 7 of \citet{Mejia16} \\
FWHM(H$\upalpha$)={\color{black}1850}~km~s$^{-1}$              	& & \\ \Tstrut\Bstrut
$\log \nu L_{\rm 5100\AA}=45.56$~erg~s$^{-1}$ 					& {\color{black}2.2$^{+0.8}_{-0.5}$} & Table 14 of \citet{Bentz13} \\
FWHM(H$\upbeta$)={\color{black}1760}~km~s$^{-1}$               	& & \\ \Tstrut\Bstrut
$\log L_{\rm H\upalpha}={\color{black}43.65}$~erg~s$^{-1}$     	& {\color{black}0.5$^{+0.3}_{-0.2}$} & Table 7 of \citet{Mejia16} \\
FWHM(H$\upalpha$)={\color{black}1850}~km~s$^{-1}$              	& & \\ \Tstrut\Bstrut
$\log L_{\rm H\upbeta}={\color{black}43.11}$~erg~s$^{-1}$      	& {\color{black}0.8$^{+0.5}_{-0.3}$} & Table 2 of \citet{Greene10} \\
FWHM(H$\upbeta$)={\color{black}1760}~km~s$^{-1}$               	& & \\ 
\hline
\end{tabular}
\parbox[]{11cm}{The $1\sigma$ error is derived from the instrinsic
  scatter rather than the errors on the best-fit parameter values.}
\end{table*}

Our estimates of the BH mass based on the ionising continuum luminosity and the
width of the broad Balmer emission lines H$\upalpha$ and H$\upbeta$ and
the Mg\,\textsc{ii} emission line give a very small range of values of
$\approx {\color{black}2.2 - 2.6} \times 10^8$~M$_{\sun}$, with an average value of
$\approx {\color{black}2.4} \times 10^8$~M$_{\sun}$. This excellent agreement between
the three estimates is remarkable, given that the relationships upon
which they are based have uncertainties of the order of $\sim 30-80\%$
at the $1\sigma$~level.

Alternative methods to estimate the BH mass, which are based on the line luminosity instead of
the ionising continuum luminosity (e.g.\ \citealt{Greene10, Mejia16}) are also considered
and we quote the results in Table~\ref{tab:bhmass}.
These estimates are in the range of $\sim {\color{black}0.5 - 0.8} \times 10^8$~M$_{\sun}$,
which are a factor of $\sim 4$ smaller than found previously.
{\color{black} 
Since we use the same broad line FWHMs in both estimates, the difference in the estimated 
masses is due to the different choice of proxy for the ionising luminosity (i.e.\ the monochromatic 
continuum luminosity or the broad line luminosity). 
We argue that it is unlikely that the difference in masses is due to an overestimation of the
3000~\AA\ and 5100~\AA\ luminosities because of unsubtracted emission 
from the host galaxy or the jet.
Firstly, for a bright AGN such as this, we expect the host galaxy emission 
to be relatively weak, as was previously noted by \cite{Yao15}.  Certainly at 3000~\AA\ in 
the rest frame the host galaxy emission will be negligible.
Second, estimates of the mass depend approximately on the square root of the (continuum or line)
luminosity.  
The two mass estimation methods would agree only if we have over-estimated the AGN contribution to $L_\mathrm{5100\si\angstrom}$ and 
$L_\mathrm{3000\si\angstrom}$
by a factor $\sim4^2$, which is highly unlikely.
If we were to believe that $L_\mathrm{5100\si\angstrom}$ and 
$L_\mathrm{3000\si\angstrom}$ are contaminated then since the  mass estimates using both luminosities are similarly high, they must contain an approximately equal fraction of contaminating emission
(i.e.\ the contaminating non-acrretion disc emission must have a spectral shape similar to that of an accretion disc); again this seems contrived.
Our SED models in the following sections also demonstrate the accretion disc emission dominates at 5100 and 3000~\AA\ (see Figures~\ref{fig:optxconv} and \ref{fig:jet-scaled}).
We have compared the two methods of BLR radius and BH mass estimations using large samples of AGN.
We find that values derived using the line luminosities are systematically smaller than those using the continuum luminosities.  
A thorough investigation of this effect is beyond the scope of this paper.
}

The BH mass was previously
estimated by \citet{Yao15}. These authors used the width of the
H$\upbeta$ broad component {\color{black}(modelled with a broad Lorentzian profile)} and the 5100~\AA~continuum luminosity and
obtained a value of $\approx 2 \times 10^8$~M$_{\sun}$, which is {\color{black}consistent with} our 
estimate{\color{black}s using the emission line FWHMs and continuum luminosities}. 
{\color{black}We adopt the mass $2\times10^8$~M$_{\sun}$ in the following.}

\section{The multiwavelength data set}
\subsection{Data quasi-simultaneous with X-shooter}
\subsubsection{NuSTAR}
We obtained an X-ray observation of PKS\,J1222$+$0413 with the \textit{Nuclear Spectroscopic Telescope Array} (\textit{NuSTAR} \citealt{Harrison13}) on 2017 June 27--18 (OBS ID: 60301018002, PI: Kynoch). 
The data reduction was performed following the method described in \cite{Landt17} (see Section 2.3.1 in that paper) and employing the latest software available (NuSTARDAS version 1.7.1, HEASOFT version 6.21 and CALDB version 20170222). 
The sum of good-time intervals after event cleaning and filtering the event files is 24.5\,ks for both FPMA and FPMB. 
The signal-to-noise ratio is 25.5 and 15.5 per module for the 3--10, and 10--79\,keV bands, respectively. 
The PSF-corrected source count rate was stable around 0.10\,s{\color{black} $^{-1}$} per module throughout the observation.

We analyzed the {\em NuSTAR} spectra using \textsc{Xspec}~\citep{Arnaud96}. FPMA and FPMB spectra were fitted simultaneously, without co-adding, with a cross-normalization factor allowed to vary during the fit. Galactic absorption was assumed to be fixed at $1.64\times 10^{20}$\,cm$^{-2}$, according to \citep{Kalberla05}. A simple power-law model already fits the data very well, with $\chi^2=91.3$ for 89 degrees of freedom (d.o.f.). The best-fit photon index is $1.33\pm0.06$, where the uncertainty is given as the 68\,\% confidence interval. Replacing the power-law continuum with a more flexible log-parabolic model ($F(E) \propto E-\alpha-\beta \log E$), we find that the curvature parameter ($\beta$) is consistent with zero within uncertainties. Fluxes in the 3--10\,keV and 10--79\,keV bands are $1.26\pm 0.05\times 10^{-12}$\,erg\,s$^{-1}$\,cm$^{-2}$ and $6.8\pm 0.5\times 10^{-12}$\,erg\,s$^{-1}$\,cm$^{-2}$, respectively.\footnote{For comparison with previous observations, the 2--10 keV flux is $1.53\pm 0.06\times 10^{-12}$\,erg\,s$^{-1}$\,cm$^{-2}$.}

\subsubsection{Swift}
A 2~ks \textit{Swift} snapshot was taken simultaneously with the \textit{NuSTAR} observation.
Data products from the X-ray telescope (XRT) were created using the xrtpipeline v0.13.2.
The source extraction region was a $47^{\prime\prime}$-radius circle centred on the source (corresponding to the 90\% encircled energy radius at 1.5~keV) and the background region was a $141^{\prime\prime}$-radius circular region offset from the source, in an area free of field sources. 
The spectra were extracted using xselect and ancilliary response files were created with xrtmkarf.
The source count rate was 0.003 counts~s$^{-1}$.

The snapshot included a UV observation with the 2600~\AA\ UVW1 filter.
Circular source and background regions of $5^{\prime\prime}$ and $30^{\prime\prime}$ radius, respectively, were analysed.
Inspecting the data with the uvotsource tool, the Vega magnitude was determined to be $16.76\pm0.04$~mag.
The tool uvot2pha was used to create \textsc{Xspec}-ready files.

\begin{figure*}
\centering
\includegraphics[width=2.1\columnwidth, keepaspectratio]{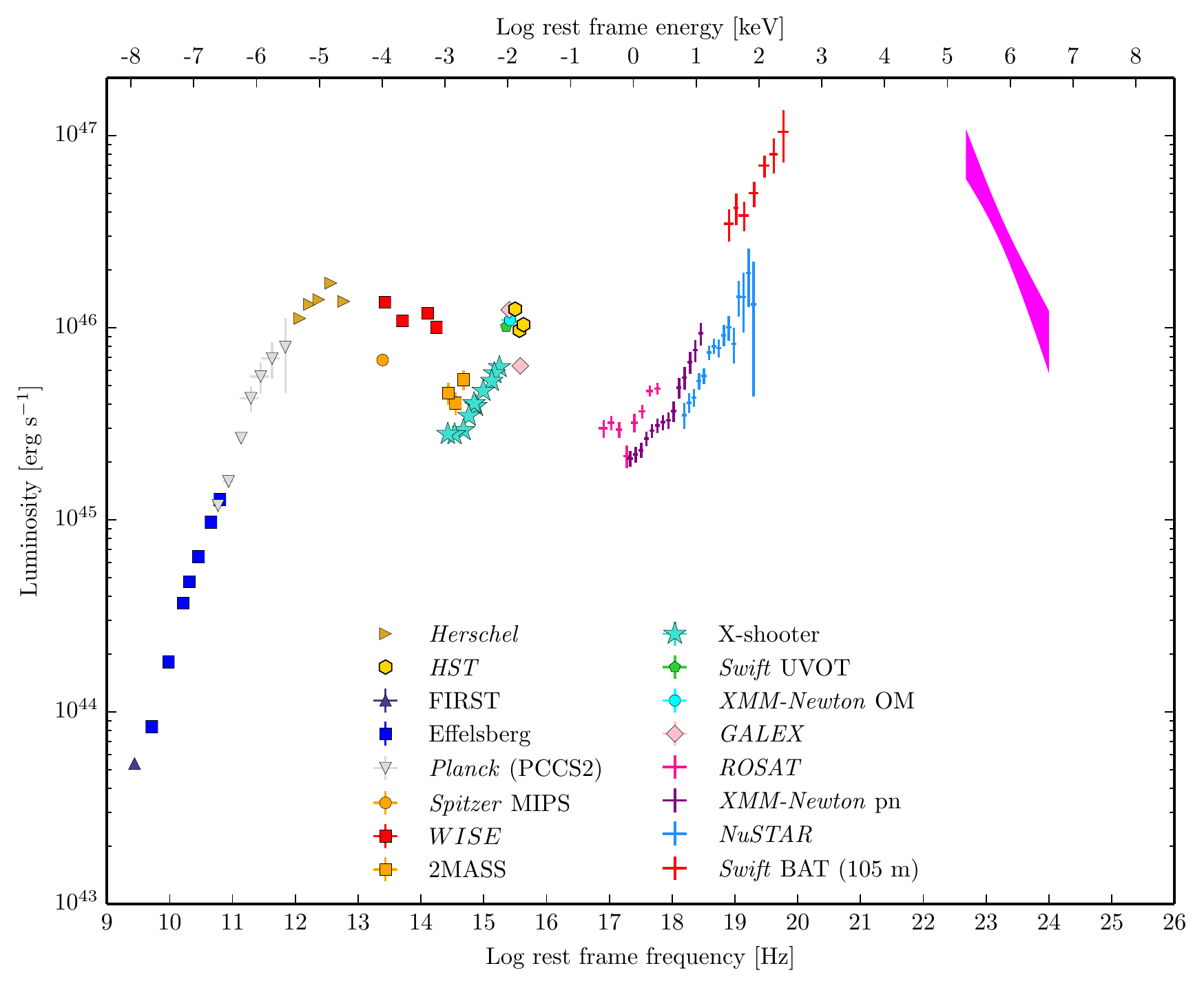}
\caption[General Caption]{The multiwavelength SED of PKS\,J1222$+$0413.
Our new data from Effelsberg (recorded between 2011 May and 2012 February), VLT X-shooter (recorded on 2017 April 3), \textit{NuSTAR} and \textit{Swift} (recorded on 2017 June 27) are shown with a \textit{Fermi} $\gamma$-ray spectrum covering the same period as our new NIR/optical/UV/X-ray data (2017 March -- July).  We also show archival data from other facilities.  
}
\label{fig:mwl}
\end{figure*}

\subsubsection{XMM-Newton}
\label{sec:xmm}
A short observation (12~ks) of PKS\,J1222$+$0413 was made by \textit{XMM-Newton} on 2006 July 12.
The observation data files (OBS ID: 0401790601) were obtained from the \textit{XMM-Newton} Science Archive and reduced using the Science Analysis System (SAS, v16.0.0).
The observation was strongly affected by background flaring, so after filtering the EPIC event files the good exposure times were 1.8, 3.8 and 3.6~ks for the pn, MOS1 and MOS2 detectors, respectively.
Circular regions of radius $40^{\prime\prime}$ and $120^{\prime\prime}$ were used to extract the source and background spectra, respectively. 
The source extraction regions were centred on the source whereas the background extraction regions were placed on a blank patch of sky on the same chip as the source.

The \textit{XMM-Newton} optical monitor (OM) observed PKS\,J1222$+$0413 using only the UVM2 filter.
The photometry for this filter was extracted using the SAS omichain and omsource tasks and standard procedures.

\begin{figure}
\centering
	\includegraphics[width=\columnwidth, keepaspectratio]{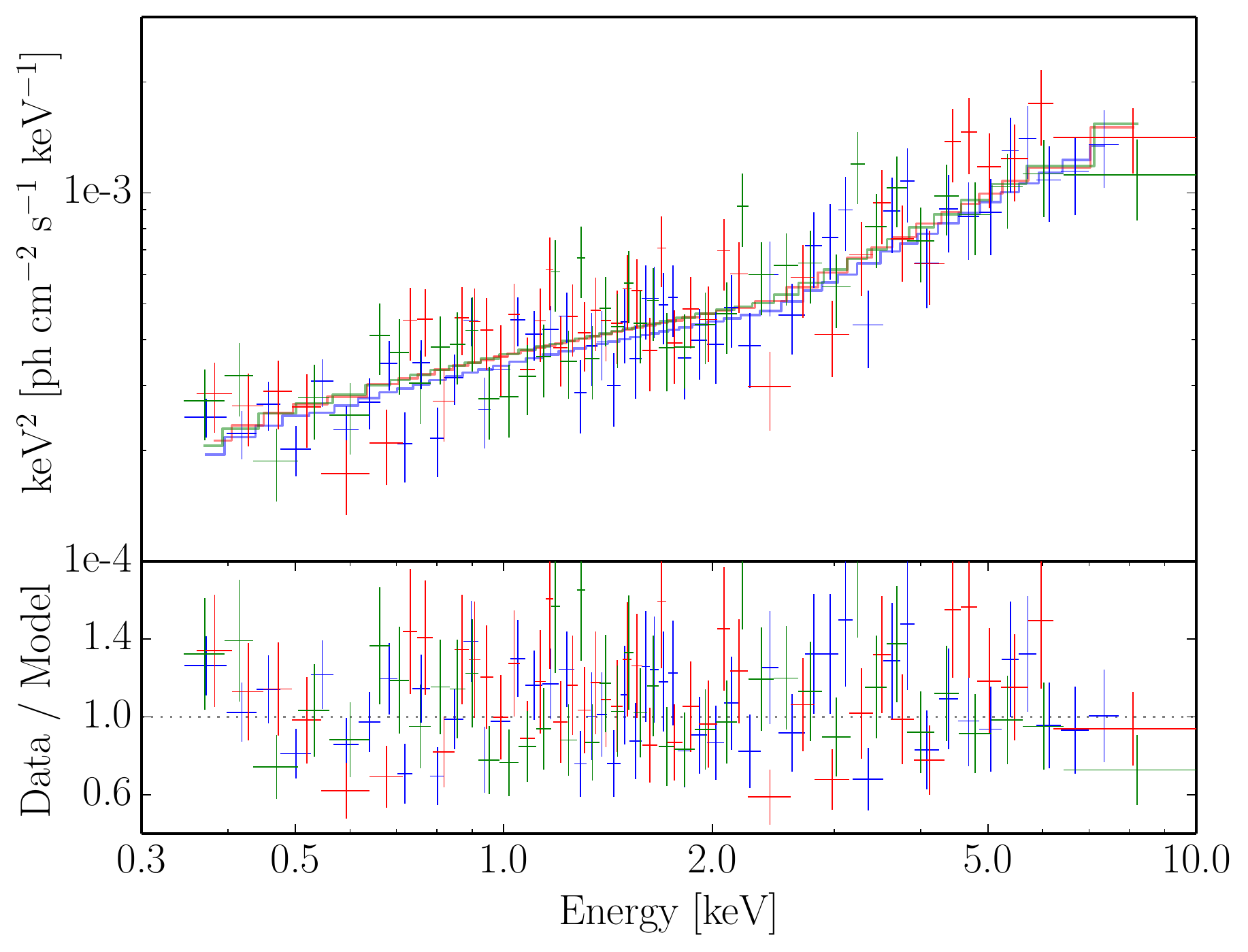}
\caption[General Caption]{\textit{XMM-Newton} EPIC X-ray spectra modelled with a broken power-law.
Data from the pn, MOS1 and MOS2 detectors are shown in blue, red and green, respectively.
The upper panel shows the data model with the detector responses folded out; the lower panel the data/model ratios.}
\label{fig:xmm}
\end{figure}

Data analysis was performed with the X-ray spectral fitting package \textsc{Xspec} v12.9.0n (\citealt{Arnaud96}).
We fitted all three EPIC (pn, MOS1 and MOS2) spectra simultaneously, allowing for a cross-normalisation factor between them; these did not vary by more than $\approx6\%$. 
In all models we include a photoelectric absorption model (\textsc{phabs}) with the Galactic neutral hydrogen absorbing column set to the value $N_\mathrm{H}=1.66\times10^{20}$~cm$^{-2}$ reported by \cite{DL90} and adopting the elemental abundances of \cite{Wilms00}.
The inclusion of an intrinsic absorber (\textsc{zphabs}) at the redshift of the source did not improve any of the fits.
A broken power-law model is a statistically significant improvement over a single power-law, with an $F$-test probability greater than 99.99\%.
A more physical model including two Comptonisation regions, \textsc{comptt} (\citealt{Titarchuk94}) plus \textsc{nthcomp} (\citealt{Zycki99}) was no improvement over the simpler broken power-law.
For the broken power-law model we determine an intrinsic flux of $(3.4\pm0.1)\times10^{-12}$~erg~s$^{-1}$~cm$^{-2}$. 
The model parameters and fit results are summarised in Table~\ref{tab:xmm} and the modelled spectra are shown in Figure~\ref{fig:xmm}. 

\begin{table}
	\centering
	\caption{Results of \textit{XMM-Newton} X-ray spectral fits}
	\begin{tabular}{lll} 
		\hline
														
		Model 							& Parameter		 							& Value \\ 
		\hline
		\Tstrut\Bstrut

		\sc{powerlaw} 					& $\Gamma$									& $1.49\pm0.03$ \\
										& norm.										& $(3.6\pm0.1)\times10^{-4}$ \\
										& $\chi^2$/d.o.f. 							& 179 / 134 = 1.34 \vspace{0.5em}\\
		
		\sc{bknpower} 						& $\Gamma_1$								& $1.65\pm0.04$ \\
										& $E_\mathrm{break}$ [keV]					& $2.4\pm0.3$ \\
								 		& $\Gamma_2$			 					& $1.1\pm0.1$ \\
										& norm.										& $(3.5\pm0.1)\times10^{-4}$ \\
										& $\chi^2$/d.o.f. 							& 151 / 132 = 1.15 \vspace{0.5em}\\
											
		\sc{comptt} +					& $kT_\mathrm{e}$ [keV]						& $0.7_{-0.1}^{+0.2}$ \\
										& $\tau$ 									& $11_{-1}^{+2}$ \\
										& norm.										& $\left(1.5_{-0.3}^{+0.5}\right)\times10^{-2}$ \\
		\sc{nthcomp}			 		& $\Gamma$			 						& $1.2\pm0.1$ \\
										& norm.										& $(1.5\pm0.4)\times10^{-4}$ \\							
										& $\chi^2$/d.o.f. 							& 151 / 131 = 1.16 \vspace{0.5em}\\
					 			
		\hline
	\end{tabular}
\parbox[]{6.5cm}{\vspace{0.2em}
Errors are quoted at the 1$\sigma$ level.  
All of the above models included a Galactic absorption component (\textsc{phabs}) with $N_\mathrm{H}=1.66\times10^{20}$~cm$^{-2}$.
The best-fit (broken power-law) model is plotted in Figure~\ref{fig:xmm}.}
\label{tab:xmm}
\end{table}

\subsubsection{Ly$\alpha$ contamination of photometry}
\label{sec:lya}
The strong Ly$\upalpha$ ($\lambda_\mathrm{rest}=1216$~\AA) line emission from the source appears at 2390~\AA\ in the observed frame; it therefore contaminates the \textit{Swift} UVOT UVW1, \textit{XMM-Newton} OM UVM2 and \textit{GALEX} NUV photometry.
Precisely correcting for this contamination in this case is difficult since we have no measurement of the Ly$\upalpha$ EW and, as noted by \cite{Elvis12}, there is no (or weak) correlation between it and the EWs of either the Balmer lines or Mg~\textsc{ii}.
However, if we assume a typical Ly$\upalpha$ EW of 60~\AA\ (in the rest frame), we can determine from the equations of \cite{Elvis12} (their section 4.1.3) that Ly$\upalpha$ contributes approximately 20 per cent of the photometric flux.

\subsubsection{Fermi}
\label{sec:fermi}
The \textit{Fermi} $\gamma$-ray source 3FGL\,J1222.4$+$0414 is associated with PKS\,J1222$+$0413. 
In the \textit{Fermi} LAT 4-year source catalogue (3FGL, \citealt{Fermi3FGL}) the $\gamma$-ray source detection significance, derived from the test statistic (TS), is $30.01\sigma$. 
In the updated  \textit{Fermi}-LAT 8yr1 (FL8Y) description of the $\gamma$-ray sky with respect to 3FGL, the source significance improved slightly ($30.77\sigma$), and the spectral shape is described as a soft power-law with an index of $\Gamma=2.87\pm0.04$, in the 100~MeV to 30~GeV energy range. 

To construct a contemporaneous SED of PKS\,J1222$+$0413, we extracted $\gamma$-ray data from the period 2017 March 20--July 11 (to include our VLT X-shooter and \textit{NuSTAR} observations) and from the period 2011 March--April (covering the \textit{HST} COS UV spectrum of March 22 and SDSS optical spectrum of March 26). Photons in a circular region of interest (RoI) of radius 10$^\circ$, centred on the position of PKS\,J1222$+$0413, were considered. 
A zenith-angle cut of 90$^\circ$ was applied. 
The analysis was done using the \textit{Fermi} Science Tools software package version v10r0p5, in combination with the PASS 8 instrument response functions (event class 128 and event type 3) corresponding to the P8R2\_SOURCE\_V6 response and the gll\_iem\_v06.fits and iso\_P8R2\_SOURCE\_V6\_v06  models for the Galactic and isotropic diffuse background, respectively. 
The spectral model of the region included all sources located within the RoI with spectral shapes and initial parameters as in the 3FGL catalogue (\citealt{Fermi3FGL}). 

\begin{figure}
\centering
	\includegraphics[width=\columnwidth, keepaspectratio]{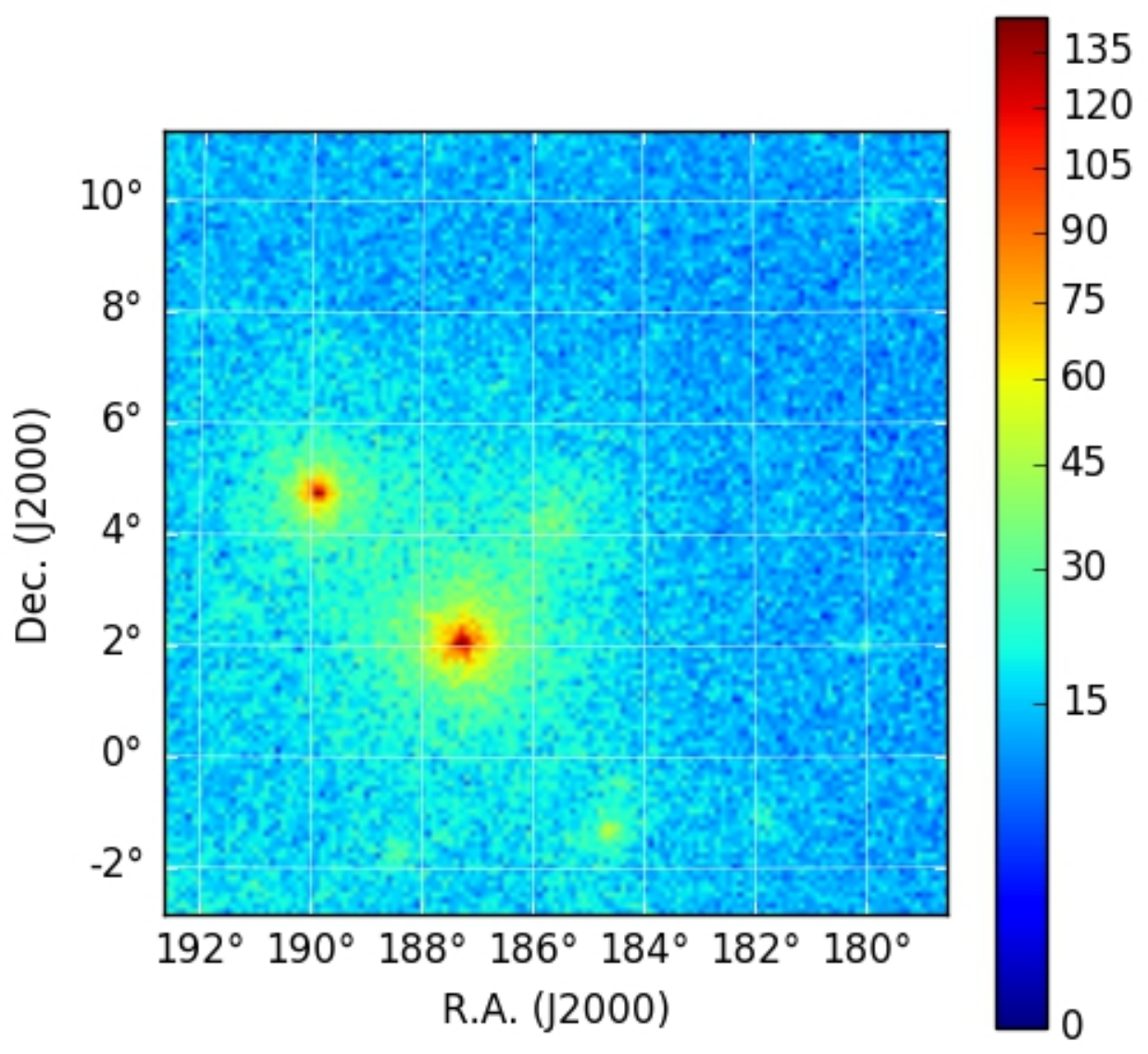}
\caption[General Caption]{\textit{Fermi} count map for the period beginning at the mission start and ending 2017 July 31.  
PKS\,J1222$+$0413 is at R.A. 185.59$^\circ$, Dec. 4.22$^\circ$.
The brightest source in the field, 3C\,273, is at R.A. 187.28$^\circ$, Dec. 2.05$^\circ$ and PKS 1237$+$049 
(3FGL~J1239.5$+$0443) is at R.A. 189.89$^\circ$, Dec. 4.72$^\circ$.}
\label{fig:fermimap}
\end{figure}

The extraction of the \textit{Fermi}-LAT data is complicated by the presence of two bright, nearby $\gamma$-ray sources in the RoI (see Figure~\ref{fig:fermimap}). 
The prototypical quasar 3C\,273 has an angular separation $\approx2.7^\circ$ from PKS\,J1222$+$0413, comparable to the \textit{Fermi} point spread function (PSF) at 1~GeV. 
Another quasar, PKS\,1237$+$049, is $\approx4.3^\circ$ from PKS\,J1222$+$0413. 
The \textit{Fermi} PSF is larger at lower energies: the 95 per cent containment radius at 100~MeV is $\geqslant 10^\circ$. 

\begin{figure}
\centering
\includegraphics[width=\columnwidth, keepaspectratio]{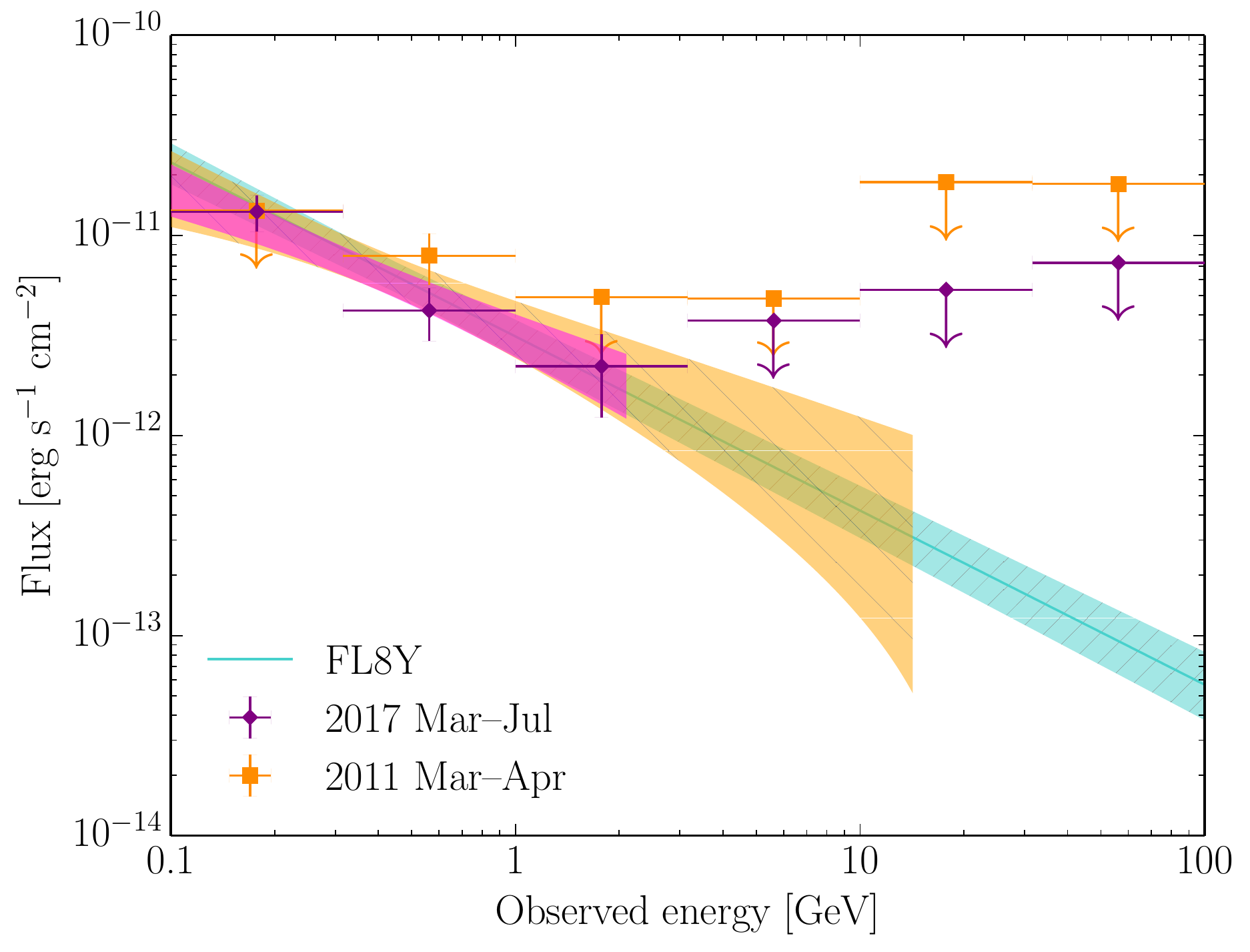}
\caption[General Caption]{\textit{Fermi} $\gamma$-ray spectra of PKS\,J1222$+$0413.
The spectrum extracted from the period 2017 March 20 to July 11 is shown in fuchsia; the corresponding model is shown as the short (0.1--2~GeV) bow-tie.
The spectrum extracted from the period 2011 March 1 to April 30 is shown in orange; the corresponding model is shown as the medium-length (0.1--14~GeV) bow-tie.
For comparison, the model spectrum taken from the \textit{Fermi}-LAT 8-year Source List (FL8Y) is shown as the long, turquoise bow-tie.}
\label{fig:fermi}
\end{figure}

In the earlier time period (2011),  PKS\,J1222$+$0413 is detected with a TS of 29.02 ($\approx5.4\sigma$), the photon index being $\Gamma=2.72\pm0.25$. 
The \textit{Fermi}-LAT flux points at two bins/decade and the model spectrum up to the highest energy photon from the source in that period are shown in Figure~\ref{fig:fermi}, although we obtain a flux detection in one low-energy bin only. 
Note that $\gamma$-ray flux upper limits at the 95 per cent confidence level are computed for bins were TS < 9.  
In the 2017 time window, PKS\,J1222$+$0413 is detected with a TS of 56.68 ($\approx7.5\sigma$), with a photon index of $\Gamma=2.74\pm0.18$. 
The last significant enrgy bin is 1--3.16~GeV. 
We thus did not attempt any study of the variability, and note that in the \textit{Fermi} All-Sky Variability Analysis (FAVA2) flare map of the region (generated in 2019 January using all available data from the mission start) that all of the flare activity is associated with other sources in the field (particularly 3C\,273) and there are no recorded flare events in the vicinity of PKS\,J1222$+$0413. 
The spectral index found in both periods is fully compatible with the entire data set (FL8Y) within statistical errors (see Figure~\ref{fig:fermi}). 

\subsection{Radio monitoring data}
\subsubsection{Effelsberg}
Observations were conducted with the 100~m Effelsberg telescope in Germany, with the secondary focus receivers at 2.64, 4.85, 8.35, 10.45, 14.6, 23.05 and 32~GHz. 
For the multiple-feed systems at 4.85, 10.45 and 32~GHz the measurements were conducted differentially thus removing most of the linear tropospheric effects. 
To correct for potential pointing offsets we have adopted the cross-scan technique; that is, register the telescope response while slewing its main beam over the expected source position. 
With our setup a complete radio SED needs roughly 40--45 minutes. 
It is therefore reasonable to assume that our observations are free of variability. 
Each pointing has been subjected to a pipeline of post-measurement corrections accounting for power losses caused by: (a) potential pointing offsets, (b) atmospheric absorption, and (c) deformations of the main telescope reflector. 
The fractional effect of each of these operations is given in Table 3 of \cite{Angelakis19}. 
The observed dataset has been finally calibrated with reference to standard sources listed in Table 3 of \cite{Angelakis15}. 
Table~\ref{tab:radio} in this paper summarises the flux densities at all observing frequencies weighted averaged over a period roughly from 2011 May 19 to 2012 February 18. 
The uncertainty in the average flux densities is limited to less that a couple of percent. 

\begin{table}
\caption{\label{tab:radio} Radio monitoring data from Effelsberg}
	\begin{tabular}{rccccc}
	\hline
	Freq.	& $\langle S \rangle$ &	$\sigma_S$ & $N$ & Start 	  & Rate    \\
	(GHz)	& (mJy)				  & (mJy) 	   &     &            & (days)  \\
	\hline
	2.64    & 663 			      & 3 	       & 9   & 2011-06-06 & 32.0    \\
	4.85    & 779 			      & 3 	       & 9   & 2011-05-19 & 34.3    \\
	8.35    & 923 			      & 5 	       & 10  & 2011-05-19 & 30.5    \\
   10.45    & 950 			      & 6 	       & 10  & 2011-05-19 & 30.5    \\
   14.60    & 920  			      & 10  	   & 10  & 2011-05-19 & 30.5    \\
   23.05    & 880  			      & 20  	   & 7   & 2011-05-19 & 45.7    \\
   32.00    & 830   			  & 20  	   & 6   & 2011-05-19 & 54.9    \\
	\hline
	\end{tabular}
\parbox[]{7.5cm}{The error-weighted mean radio flux densities and their $1\sigma$ errors, the number of detections $N$ in the observation period starting on the given date and ending on 2012-02-18.}
\end{table}

\subsection{Additional archival data}
\subsubsection{FIRST}
The Jansky Very Large Array (JVLA) Faint Images of the Radio Sky at Twenty centimetres (FIRST; \citealt{FIRST2}) project surveyed 10575 square degrees of the sky at 1.4~GHz between 1993 and 2011.
Querying the source catalogue, we find a radio source 0.5$^{\prime\prime}$ from the optical coordinates of PKS\,J1222$+$0413, which is the only radio source within a search radius of 30$^{\prime\prime}$.
The integrated flux density was reported to be 800.63~mJy and we adopt an error of 5 per cent, representative of the systematic uncertaintiy for bright sources (\citealt{FIRST1}). 
The catalogue fluxes are measured in coadded images from multiple observations; the mean observation date is given as 2001 March 14 with an RMS of $\approx4$~days.

\subsubsection{Planck}
ESA's \textit{Planck} space observatory (\citealt{Tauber10}) operated from 2009 July, mapping the sky from high-frequency radio to far-infrared wavelengths.
Its High-Frequency Instrument depleted its supply of liquid helium coolant in 2012 January, after which only the Low-Frequency Instrument (recording the 30, 44 and 70~GHz bands) was operable until the science mission ended in 2013 October.

We perform a cone search using the default radius equal to the beam FWHM at each frequency (see \citealt{PCCS2}).
In Table~\ref{tab:data} we give the fluxes derived from the quoted catalogue flux densities which were calculated using the Gaussian fitting method.
The source is not detected at 545 or 857~GHz either in the PCCS2 catalogue or the lower-reliability PCCS2E catalogue.

\subsubsection{Herschel}
The \textit{Herschel Space Observatory} (\citealt{Pilbratt10}) operated contemporaneously with \textit{Planck}, from 2009 July to 2013 April.
\textit{Herschel} carried three science instruments: the Photodetecting Array Camera and Spectrometer (PACS, covering 55--220~$\upmu$m), the Spectral and Photometric Imaging Receiver (SPIRE, containing a low-resolution spectrometer covering 194--672~$\upmu$m and a photometer with three bands centred on 250, 350 and 500~$\upmu$m) and the Heterodyne Instrument for the Far Infrared (HIFI, covering 157--210 and 236--615~$\upmu$m).
We searched the PACS and SPIRE point source catalogues (\citealt{Herschel-1}; \citealt{Herschel-2}) and obtained the data listed in Table~\ref{tab:data}. 

\subsubsection{Spitzer}
PKS\,J1222$+$0413 has been observed by the \textit{Spitzer Space Telescope} (\citealt{Spitzer}).
The source was detected by the 24~$\upmu$m array of the the Multiband Imaging Photometer for \textit{Spitzer} (MIPS) during an observation made on 28 January 2005. 
The source flux density is taken from the Spitzer Enhanced Imaging Products\footnote{\url{http://irsa.ipac.caltech.edu/data/SPITZER/Enhanced/SEIP/}} (SEIP) source list.


\subsubsection{WISE}
The \textit{Wide-field Infrared Survey Explorer} (\textit{WISE}; \citealt{WISE10}) makes photometric observations in four bands: W1 (3.4~$\upmu$m), W2 (4.6~$\upmu$m), W3 (12~$\upmu$m) and  W4 (22~$\upmu$m).
For each band we have calculated the flux from the instrumental profile-fit photometry magnitude listed in the AllWISE catalogue\footnote{\url{http://irsa.ipac.caltech.edu/}}.
The magnitudes recorded for the bands were W1: $12.802\pm0.023$; W2: $11.634\pm0.023$; W3: $8.901\pm0.034$ and W4: $6.513\pm0.080$;
the photometric quality of each band was reported to be of the highest quality. 

\begin{figure}
\centering
\includegraphics[width=\columnwidth, keepaspectratio]{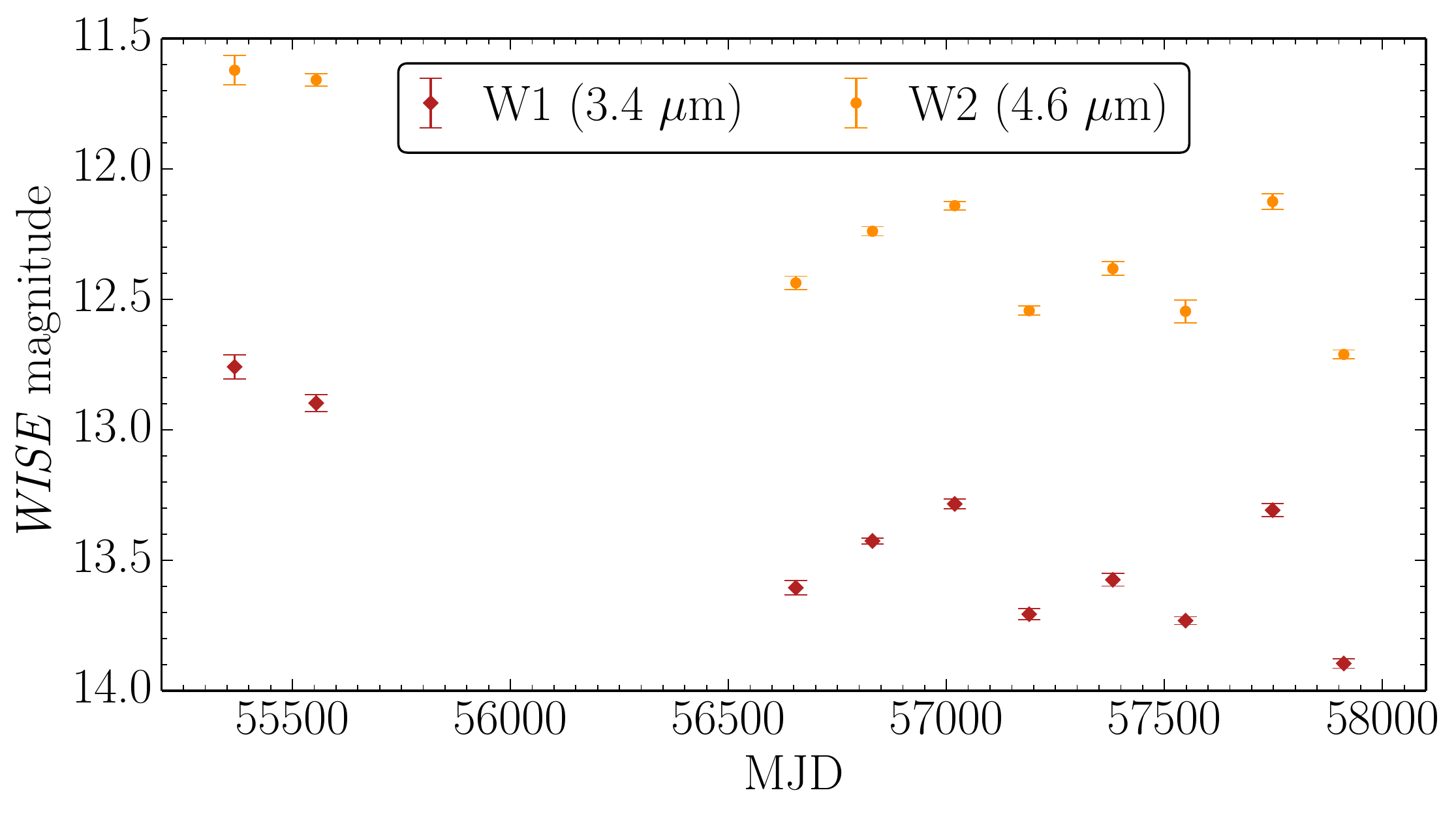}
\caption[General Caption]{The \textit{WISE} and \textit{NEOWISE} infrared lightcurve of PKS\,J1222$+$0413, showing observations made between 2010 June and 2017 June.}
\label{fig:wiselc}
\end{figure}

The \textit{WISE} telescope was placed into hibernation in 2011 February, following the depletion of its coolant.  
It was reactivated in 2013 September to begin the \textit{NEOWISE} mission (\citealt{Mainzer14}) using only the the two shortest-wavelength filters.
Infrared photometry spanning several years is available for these filters, which sample the hot dust in our source.
In Figure~\ref{fig:wiselc} we show the infrared lightcurves.
Several exposures were taken on each pass of the telescope (with a cadence of roughly six months), which we have binned into the ten single-epoch measurements shown.
For each epoch we have calculated the mean magnitude and its standard error from the individual exposures, after applying a $3\sigma$ clipping algorithm to remove anomalous data.
The peak-to-peak magnitude change (from the first \textit{WISE} observation in mid-2010 to the latest \textit{NEOWISE} observation in mid-2017) is $\approx1.2$~mag, and $\approx0.6$~mag changes are seen in the \textit{NEOWISE} period.

\subsubsection{2MASS}
PKS\,J1222$+$0413 was detected as part of the Two Micron All Sky Survey (2MASS, \citealt{Skrutskie06}), conducted between 1997 and 2001.
The J, H and K$_\mathrm{s}$ profile-fit instrumental magnitudes recorded in the 2MASS All-Sky Point Source Catalog (PSC) were $16.348\pm0.133$, $15.849\pm0.146$ and $14.969\pm0.140$, respectively.

\subsubsection{Sloan}
\label{sec:sdss}
Two optical spectra of the source have been recorded as part of the Sloan Digital Sky Survey (SDSS).
The first was taken on 12 February 2008 with the original SDSS spectrograph; the second was taken on 26 March 2011 with the Baryon Oscillation Spectroscopic Survey (BOSS) spectrograph, allowing a greater wavelength coverage.
The two spectra are compared with our recent X-shooter spectrum in Figure~\ref{fig:xsh}.
It can be seen that the 2011 spectrum is redder than that of 2008.
\cite{Margala16} noted that in the spectra of BOSS quasars, the flux densities were overestimated by $\sim19$~per~cent at 3300~\AA\ 
and underestimated it by $\sim24$~per~cent at 1~$\upmu$m.
Given the $\sim1^{\prime\prime}$ seeing at 500~nm during our X-shooter observation, slit losses are likely minimal in the optical and UV, for which we used $1.2^{\prime\prime}$ and $1.3^{\prime\prime}$ width slits, respectively.

In the observed frame, the flux density at 4000~\AA\ (7000~\AA) is approximately 30 (25) per cent greater in the 2008 SDDS spectrum
than in the 2017 X-shooter spectrum, and the flux densities of the two SDSS spectra are similar at 4000~\AA.
 

\subsubsection{GALEX}
The \textit{Galaxy Evolution Explorer} (\textit{GALEX}; \citealt{Galex05}), an ultraviolet space telescope, detected the source in both its near-ultraviolet (NUV) and far-ultraviolet (FUV) bands.  The FUV flux is severely absorbed because the filter (1340--1806~\AA) covers a spectral region blueward of the Lyman break of a Lyman limit system at 1793~\AA\ (observed).

\subsubsection{Hubble Space Telescope}
\label{sec:hubble}
\begin{figure}
\centering
	\includegraphics[width=\columnwidth, keepaspectratio]{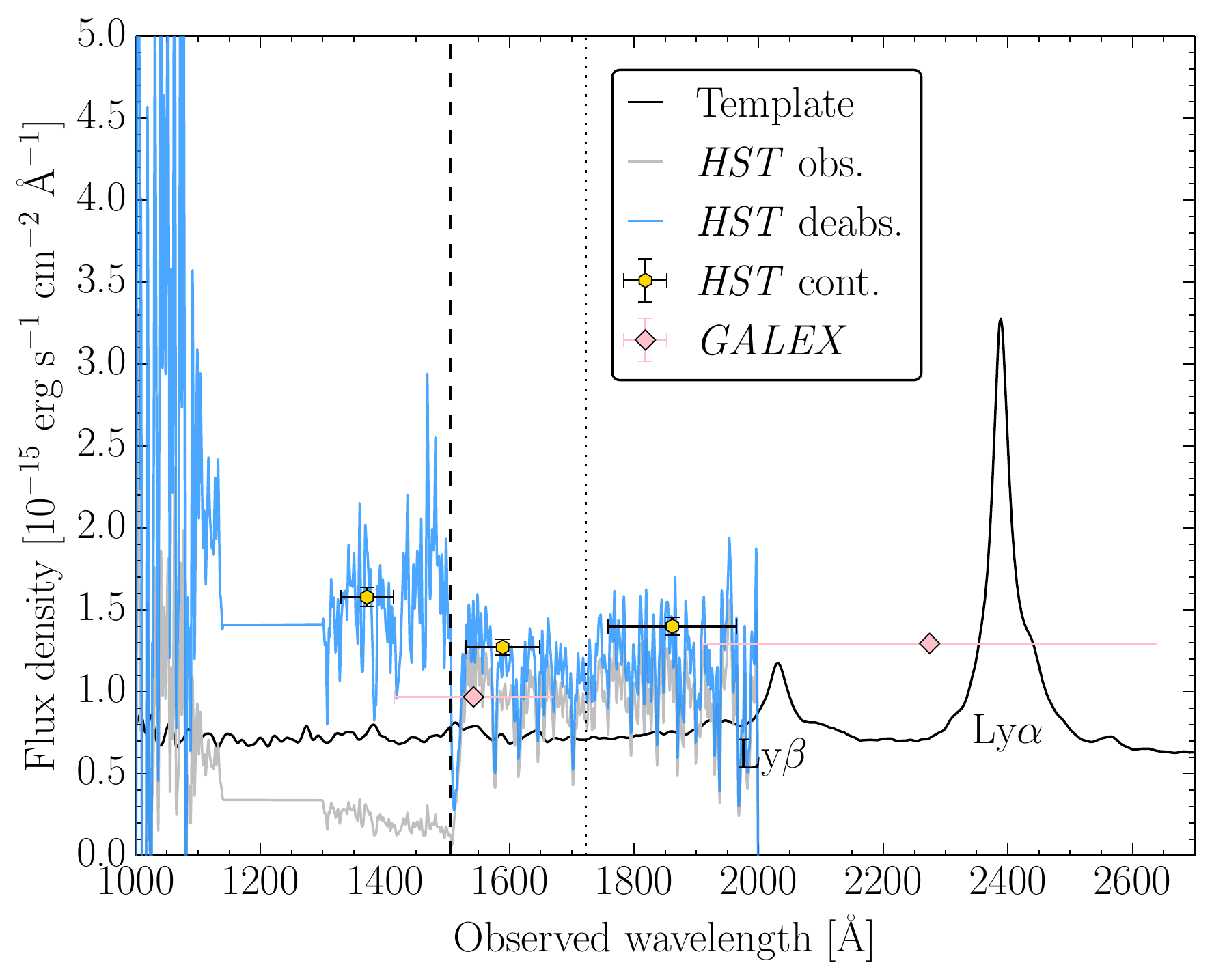}
\caption[General Caption]{The \textit{HST} COS spectrum of PKS\,J1222$+$0413.  
The observed spectrum is shown in grey.  Overplotted in blue is the spectrum corrected for Galactic reddening and the Lyman continuum absorption of a Lyman limit system (LLS) at $z_\mathrm{LLS}=0.65$. 
The dashed vertical line indicates the Lyman break of the LLS; the dotted vertical line indicates the Lyman break of an intervening partial LLS at $z_\mathrm{pLLS}=0.89$.
\textit{GALEX} NUV and FUV photometry (corrected for Galactic reddening) are shown as pink diamonds.
For reference, the composite spectrum of \cite{Telfer02} is shown in black.  
This template has been scaled to approximately match the continuum flux level of the 2011 SDSS spectrum.}
\label{fig:hst}
\end{figure}

\cite{Wotta16} studied absorbing systems on sight lines towards 61 AGN, including PKS\,J1222$+$0413.
As part of this study they obtained a short (900~s) exposure of the UV spectrum of the source using the G140L (900--2150~\AA) filter of the \textit{HST} Cosmic Origins Spectrograph (COS).  
This snapshot spectrum was recorded on 2011 March 22. 
They identified a Lyman limit system (LLS) with $\log(N_\mathrm{H})=17.55\pm0.10$~cm$^{-2}$ at redshift $z_\mathrm{LLS}=0.6547$.
The $N_\mathrm{H}$ was determined by assessing the suppression of continuum flux density relative to the composite quasar template of \cite{Telfer02}\footnote{The template shown in Figure~\ref{fig:hst} is the composite of the radio loud subset of sources.  It is constructed from 205 spectra of 107 objects, all with $z>0.33$ and with $\langle z \rangle = 1.00$.}, scaled to match the unabsorbed region of the spectrum.
We correct the observed spectrum for the Lyman continuum absorption by multiplying the fluxes by a factor $e^{\tau_\lambda}$ where {\color{black} the optical depth}
\begin{equation}
\tau_\lambda = \sigma{\color{black}_0} \left(\frac{\lambda}{\lambda_\mathrm{LLS}}\right)^3 N_\mathrm{H}~~\mathrm{for}~~\lambda\leqslant\lambda_\mathrm{LLS},
\label{eqn:tau}
\end{equation}
{\color{black}the absorption cross-section at the Lyman limit $\sigma_0=6.3\times10^{-18}$~cm$^2$ and} $\lambda_\mathrm{LLS} = (1 + z_\mathrm{LLS})\times912$~\AA\ {\color{black} (\citealt{Shull17}; \citealt{Paresce84})}.
The observed and corrected spectra are shown in Figure~\ref{fig:hst}.

To estimate the AGN continuum, we select several regions free from narrow absorption lines and calculate the mean flux density in each of these and then bin into three wider regions, shown in Figure~\ref{fig:hst}.
It should be noted that the estimate of the column density of the LLS is dependent on the shape of the composite template used to approximate the intrinsic continuum.  
Without an independent measure of the column density, our `recovery' of the PKS\,J1222$+$0413 AGN continuum is therefore also dependent on the shape of the template, so the flux density determined at 1370~\AA\ should be treated with some caution.
The continuum flux density at 1590~\AA\ is perhaps slightly underestimated because of the intervening partial LLS at $z_\mathrm{pLLS}=0.89$, the column density of which is undetermined.
The continuum flux density at 1860~\AA\ is more reliable although all three estimates are hampered by the limited signal-to-noise of the spectrum. 
The spectrum blueward of $\approx1100$~\AA\ is too noisy to be of use.

\subsubsection{ROSAT}
An 8~ks pointed observation of PKS\,
J1222$+$0413 was made by the \textit{R\"{o}ntgensatellit} (\textit{ROSAT}; \citealt{Voges99}) on 24 December 1992.
The archival data was reduced using HEAsoft \textsc{xselect} v2.4d.
The source extraction region was chosen to be a 120$^{\prime\prime}$-radius circle centred on the source.
The background region was a 360$^{\prime\prime}$-radius circle offset on a blank patch of sky, on the same chip as the source.
The 0.1--2.0~keV source count rate was $\approx0.15$~counts~s$^{-1}$.  
The source spectrum was rebinned using the \textsc{grppha} tool to contain a minimum of 20 counts per bin.

The best fitting model was a simple power-law with $\Gamma=1.72\pm0.04$ and normalisation $(5.2\pm0.2)\times10^{-4}$, giving $\chi^2_\nu=90/57=1.58$.
The best fit $N_\mathrm{H}=(7\pm2)\times10^{-3}$, lower than the \cite{DL90} value but improving the fit by only $\Delta\chi^2=9$ for one additional free parameter.

\subsubsection{Swift BAT}
In the broad-band SED we also include data from the \textit{Swift} Burst Alert Telescope (BAT) 105-month all-sky hard X-Ray survey (\citealt{Oh18}) spanning the period 2004 December -- 2013 August.
The time-averaged spectrum and response file were obtained from the online archive.
The spectrum was well-fit in \textsc{Xspec} with a single power-law of index $\Gamma=1.5\pm0.1$ and normalisation $\left(9^{+5}_{-3}\right)\times10^{-4}$, giving $\chi^2_\nu=0.32$.


\begin{table*}
	\centering
	\caption{The multiwavelength data set}
	\label{tab:data}
	\begin{tabular}{clll *{1}{S[table-format=-1.3]} llc}
		\hline
		Q 			& Band				& Instrument						& Observation date		& {$\log(\nu_\mathrm{obs})$} & Flux 					& Luminosity					& Ref.\ \\
					& 					& (Survey)							& (D/M/Y or M/Y)		& [Hz]				& [$10^{-14}$~erg/s/cm$^2$]		& [$10^{43}$~erg/s]				&  \\ 
		\hline
					& Radio				& (FIRST)							& 03/01					& 9.15		 		& $1.12\pm0.06$					& $5.4\pm0.3$ 					& [1] \\
					& Radio				& Effelsberg						& 06/11--02/12          & 9.42		 		& $1.750\pm0.0008$				& $8.40\pm0.04$ 				& [7] \\
					& Radio				& Effelsberg						& 05/11--02/12 			& 9.69		 		& $3.80\pm0.01$					& $18.3\pm0.07$ 				& [7] \\
					& Radio				& Effelsberg						& 05/11--02/12          & 9.92		 		& $7.71\pm0.04$					& $37.0\pm0.2$ 					& [7] \\
					& Radio				& Effelsberg						& 05/11--02/12 			& 10.02		 		& $9.93\pm0.06$					& $47.7\pm0.3$ 					& [7] \\
					& Radio				& Effelsberg						& 05/11--02/12          & 10.16		 		& $13.4\pm0.2$					& $64.5\pm0.7$ 					& [7] \\
					& Radio				& Effelsberg						& 05/11--02/12 			& 10.36			 	& $20.3\pm0.5$					& $97\pm2$ 						& [7] \\
					& Radio				& \textit{Planck}					& 08/09--08/13			& 10.48		 		& $25\pm2$			 			& $119\pm9$					& [2] \\
					& Radio				& Effelsberg						& 05/11--02/12 			& 10.51			 	& $26.6\pm0.6$					& $128\pm3$ 					& [7] \\
					& Radio				& \textit{Planck}					& 08/09--08/13			& 10.64		 		& $33\pm2$			 			& $158\pm11$					& [2] \\
					& Radio				& \textit{Planck}					& 08/09--01/12			& 10.85		 		& $55\pm3$			 			& $266\pm16$					& [2] \\
					& Radio				& \textit{Planck}					& 08/09--01/12			& 11.00		 		& $89\pm14$			 			& $429\pm66$					& [2] \\
					& Radio				& \textit{Planck}					& 08/09--01/12			& 11.16		 		& $116\pm22$			 		& $560\pm100$					& [2] \\
					& Radio				& \textit{Planck}					& 08/09--01/12			& 11.34 	 		& $144\pm31$			 		& $690\pm150$					& [2] \\
					& Radio				& \textit{Planck}					& 08/09--01/12			& 11.55		 		& $165\pm70$			 		& $790\pm340$					& [2] \\
					& Far-IR			& \textit{Herschel}					& 03/07/10			 	& 11.78				& $233\pm13$			 		& $1117\pm60$					& [3] \\
					& Far-IR			& \textit{Herschel}					& 03/07/10			 	& 11.93			 	& $276\pm14$			 		& $1322\pm67$					& [3] \\
					& Far-IR			& \textit{Herschel}					& 03/07/10			 	& 12.08		 		& $292\pm17$			 		& $1399\pm81$					& [3] \\
					& Far-IR			& \textit{Herschel}					& 03/07/10 			 	& 12.27		 		& $356\pm41$			 		& $1706\pm200$					& [3] \\
					& Far-IR			& \textit{Herschel}					& 03/07/10			 	& 12.48		 		& $285\pm21$			 		& $1369\pm100$					& [3] \\
		  
					& Far-IR			& \textit{Spitzer} MIPS				& 28/01/05				& 13.11 		 	& $142\pm1$ 					& $679\pm6$ 					& [4] \\
					& Mid-IR			& \textit{WISE} 					& 18--22/06/10			& 13.13				& $283\pm21$					& $1358\pm100$ 					& [5] \\
					& Mid-IR			& \textit{WISE} 					& 18--22/06/10			& 13.41				& $226\pm7$ 					& $1085\pm34$ 					& [5] \\
					& Mid-IR			& \textit{WISE} 					& 06/10--12/10			& 13.81		 		& $248\pm5$						& $1191\pm25$					& [5] \\
					& Mid-IR			& \textit{WISE} 					& 06/10--12/10			& 13.95		 		& $210\pm4$						& $1005\pm21$					& [5] \\
					& Near-IR			& (2MASS)							& 25/02/00				& 14.14 		 	& $95\pm12$						& $457\pm59$ 					& [6] \\
					& Near-IR			& (2MASS)							& 25/02/00				& 14.26		 		& $84\pm11$ 					& $405\pm55$ 					& [6] \\
	$\checkmark$	& Near-IR			& VLT X-shooter						& 03/04/17				& 14.27		 		& $57.3\pm0.5$					& $275\pm2$						& [7] \\
					& Near-IR			& (2MASS)							& 25/02/00				& 14.39		 		& $112\pm14$ 					& $536\pm65$ 					& [6] \\
	$\checkmark$	& Optical			& VLT X-shooter						& 03/04/17				& 14.48		 		& $74.2\pm1.1$					& $363\pm5$ 					& [7] \\
					& Optical			& (SDSS)							& 12/02/08				& 14.48				& $124\pm8$						& $621\pm40$					& [8] \\
	$\dagger$		& Optical			& (SDSS)							& 26/03/11				& 14.48				& $122\pm2$						& $596\pm10$					& [9] \\
	$\dagger$		& UV    			& (SDSS)							& 26/03/11				& 14.71				& $140\pm7$						& $699\pm35$					& [9] \\
	$\checkmark$	& UV				& VLT X-shooter						& 03/04/17 				& 14.71		 		& $91.2\pm0.6$					& $513\pm3$						& [7] \\
	$\checkmark$	& UV				& \textit{Swift} UVOT				& 27/06/17				& 15.06	 			& $211\pm8$						& $1011\pm37$					& [7] \\
					& UV				& \textit{XMM-Newton} OM			& 12/07/06 				& 15.13		 		& $228\pm4$						& $1094\pm19$ 					& [7] \\
					& UV				& \textit{GALEX}					& 31/03/04 				& 15.12				& $250\pm8$						& $1199\pm38$ 					& [10] \\
	$\dagger$		& UV				& \textit{HST} COS					& 22/03/11				& 15.21		 		& $220\pm20$					& $1240\pm90$					& [7] \\
	$\dagger$		& UV				& \textit{HST} COS					& 22/03/11				& 15.27		 		& $180\pm20$					& $1000\pm100$					& [7] \\
					& UV				& \textit{GALEX}					& 31/03/04				& 15.29		 		& $135\pm16$					& $649\pm77$					& [10] \\
	$\dagger$		& UV				& \textit{HST} COS					& 22/03/11				& 15.35			 	& $30\pm3$						& $1040\pm20$					& [7] \\
					& X-ray				& \textit{ROSAT} 					& 24/12/92 				& 17.35			 	& $100\pm9$						& $500\pm40$ 					& [7] \\
	$\checkmark$	& X-ray				& \textit{Swift} XRT				& 27/06/17				& 17.37		 		& $46\pm12$ 					& $230\pm60$ 					& [7] \\
					& X-ray				& \textit{XMM-Newton} EPIC			& 12/07/06 				& 18.08		 		& $150\pm8$ 					& $720\pm40$ 					& [7] \\
	$\checkmark$	& X-ray				& \textit{NuSTAR} 					& 27/06/17				& 18.86		 		& $220\pm10$					& $1060\pm50$					& [7] \\
					& X-ray				& \textit{Swift} BAT				& 12/04--09/10			& 19.18				& $1450\pm200$					& $6960\pm960$					& [11] \\
	$\checkmark$	& $\gamma$-ray		& \textit{Fermi} LAT				& 03/17--06/17			& 22.63	 			& $1300\pm270$					& $6300\pm1300$					& [7] \\
	$\checkmark$	& $\gamma$-ray		& \textit{Fermi} LAT				& 03/17--06/17			& 23.13	 			& $420\pm120$					& $2020\pm600$ 					& [7] \\
	$\checkmark$	& $\gamma$-ray		& \textit{Fermi} LAT				& 03/17--06/17			& 23.63				& $220\pm100$					& $1100\pm470$ 					& [7] \\
	$\checkmark$	& $\gamma$-ray		& \textit{Fermi} LAT				& 03/17--06/17			& 24.13		 		& $<370$						& $<1800$ 						& [7] \\
	$\checkmark$	& $\gamma$-ray		& \textit{Fermi} LAT				& 03/17--06/17			& 24.63				& $<530$						& $<2600$ 						& [7] \\
	$\checkmark$	& $\gamma$-ray		& \textit{Fermi} LAT				& 03/17--06/17			& 25.13	 			& $<730$						& $<3500$ 						& [7] \\
	\hline
	\end{tabular}
	\parbox[]{15cm}{\vspace{0.2em} References: [1] FIRST catalogue, \cite{FIRST2}; [2] \textit{Planck} Second Point Source Catalog, \cite{PCCS2}; [3] \textit{Herschel} point source catalogues, \cite{Herschel-1}; [4] \textit{Spitzer} SEIP Source List, \cite{Spitzer}; [5] \textit{WISE} AllWISE Source Catalog, \cite{WISE10}; [6] Two Micron All-Sky Survey, \cite{Skrutskie06}; [7] this work; [8] Sloan Digital Sky Survey DR7, \cite{SDSS-DR7}; [9] Sloan Digital Sky Survey DR9, \cite{SDSS-DR9}; [10] \textit{GALEX} Data Release GR6, \cite{Galex05}; [11] \textit{Swift} BAT 105-month all-sky hard X-Ray survey, \cite{Oh18}.
	The `Q' flag indicates our quasi-simultaneous data ($\checkmark$ for data used in our SED fitting of Section~\ref{sec:optxconv} and $\dagger$ for optical/UV data used in Section~\ref{sec:hubble}).}
\end{table*}


\section{The origin of the $\gamma$-ray emission}
{\color{black}
\subsection{Determining the external photon field}
Here, we determine the parameters of the ambient photon field with which the jet interacts to produce the observed high-energy emission.
In the following we use our multiwavelength data to determine the size scales and luminosities of the accretion disc and its corona, the broad emission line region and the hot dusty torus.
These are applied in our jet model which is described in Section~\ref{sec:jetmodel}.
}
\subsubsection{The accretion flow}
\label{sec:optxconv}
\begin{figure}
	\begin{tabular}{c}
	\includegraphics[width=\columnwidth, keepaspectratio]{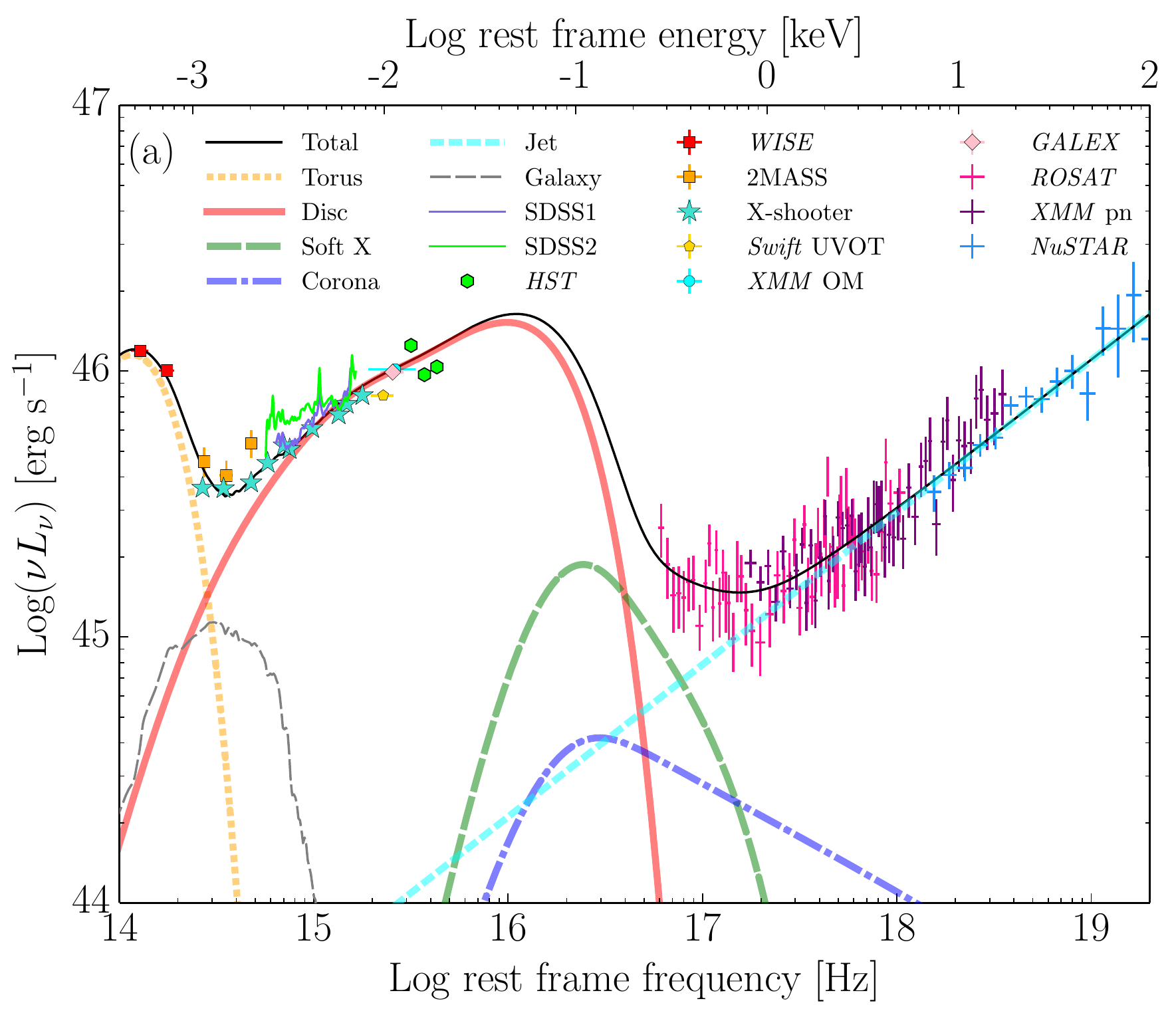}     \\
	\includegraphics[width=\columnwidth, keepaspectratio]{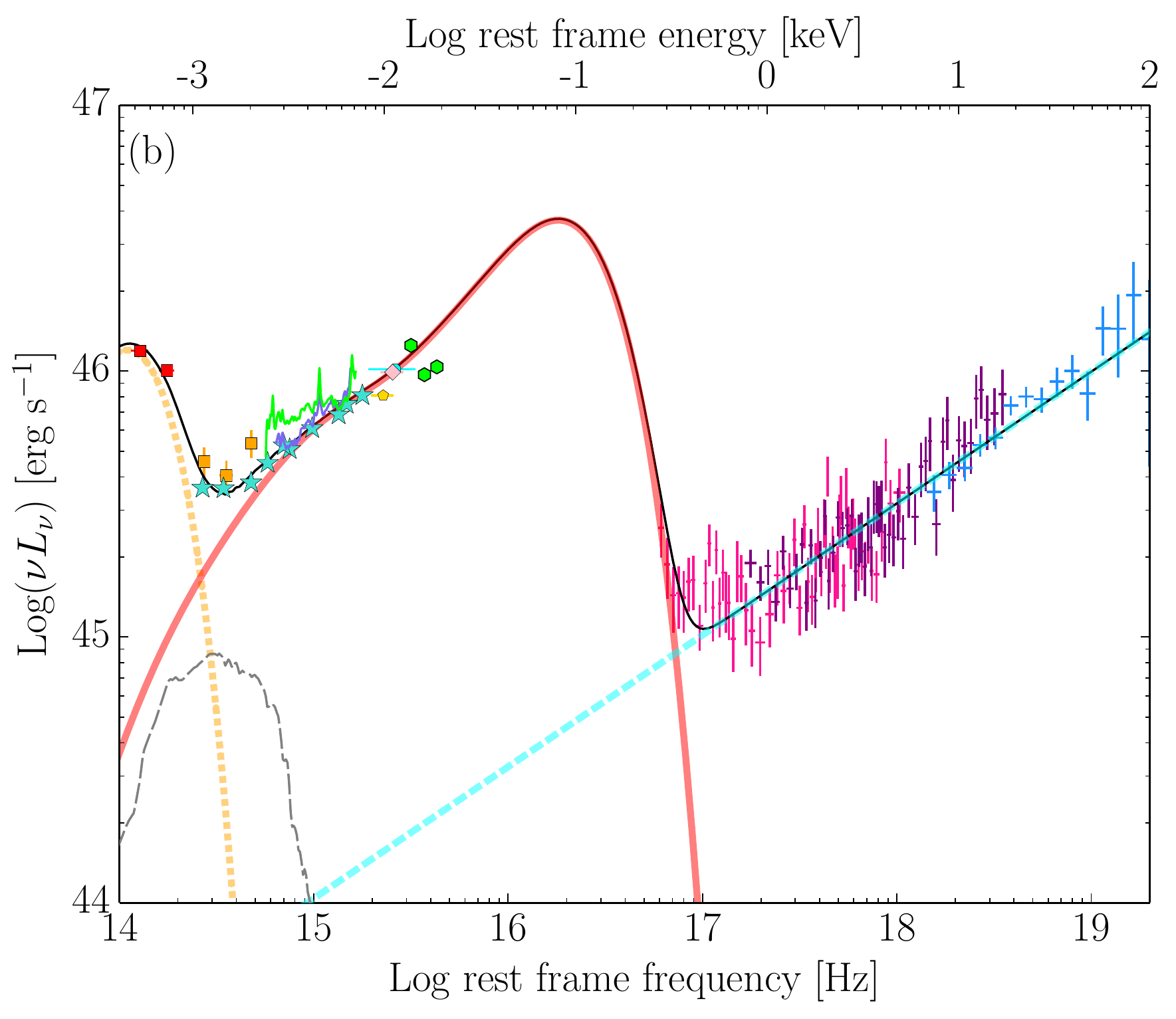} \\
	\end{tabular}
\caption{Accretion flow models applied to the NIR to X-ray SED, for different BH spins: (a) $a_\star=0$; (b) $a_\star=0.8$.}
\label{fig:optxconv}
\end{figure}

To model the infrared-to-X-ray SED we have scaled up the X-shooter fluxes by a factor 1.3 to match the level of the SDSS spectra in the UV (see \S~\ref{sec:sdss} and Figure~\ref{fig:optxconv}).
The \textit{Swift} UVOT, \textit{XMM} OM and \textit{GALEX} NUV photometry have been scaled by a factor 0.8 to account for Ly$\upalpha$ contamination (see \S~\ref{sec:lya}).
The \textit{XMM} pn and \textit{ROSAT} X-ray spectra have been scaled by factors 0.8 and 0.5, respectively, to match the flux level of our joint \textit{NuSTAR}-\textit{Swift} XRT observation.
Note that the 2011 SDSS spectrum and \textit{HST} COS spectrum (shown in lime green in Figure~\ref{fig:optxconv}) were recorded three days apart.

We employ the energy-conserving accretion flow model, \textsc{optxconv}, of \cite{Done13} and perform a simple test to determine whether the spin of the BH can be constrained by our data.
In the \textsc{optxconv} model, a colour-temperature corrected accretion disc spectrum is produced from the accretion disc outer radius (which we set to the self-gravity radius) down to $R_\mathrm{cor}$,
inside of which the accretion power is divided between warm and hot Comptonisation regions which are the origin of the observed soft X-ray excess and coronal power-law emission, respectively.
As the BH spin increases, the innermost stable circular orbit moves closer to the BH and (for fixed $R_\mathrm{cor}$) this increases the emission from the Comptonisation regions.
In this test we fix several parameters to those typical of a NLS1; namely, the warm Comptonisation region electron temperature $kT_\mathrm{e}{\color{black}=0.2}${\color{black}~keV} and optical depth $\tau{\color{black}=15}$, the photon index of the power-law tail $\Gamma{\color{black}_\mathrm{PL}=2.4}$ and the ratio of power-law to soft excess power $f_\mathrm{PL}{\color{black}=0.3}$.
{\color{black}The outer accretion disc radius is fixed at 1000~$R_\mathrm{g}$.}

As well as the accretion flow, we model the hot dust emission as simple blackbody, a scaled host galaxy template\footnote{We use the 5~Gyr-old elliptical galaxy template of \citealt{Polletta07}, scaled so that the total model fits the data at $\approx1~\upmu$m.} and add a power-law with $\Gamma=1.4$ to model the putative jet emission at hard X-ray energies.
The hot dust temperature and luminosity is well determined by the \textit{WISE} W1 and W2 points.
The hot dust temperature is well-determined from the \textit{WISE} W1 and W2 points and {\color{black} changes little} between fits.
As can be seen in Figure~\ref{fig:optxconv}, our data do not strongly discriminate between zero- and high-spin models.
{\color{black}However, the high-spin model slightly overpredicts the soft X-ray flux (although the modelled \textit{ROSAT} spectrum has been scaled down by 20 per cent to match the flux level of the \textit{Swift} XRT data) and cannot accommodate soft Comptonisation or coronal components.
We note that this model conserves energy between the outer and inner accretion flow, producing the optical and UV/X-ray emission, respectively.
The UV/X-ray flux would be lower if power were lost between outer and inner radii  as the result of an accretion disc wind which could be present if the accretion rate were super-Eddington as in the high-spin model.
We cannot therefore rule out a high BH spin.
Overall, the zero-spin model is a better representation of the data and the parameters of the accretion flow components are then typical of a NLS1.}


\begin{table*}
\small
\caption{\label{tab:optx} Results from spectral fits to the deabsorbed IR to hard X-ray SED \vspace{-2em}}
\begin{center}
\begin{tabular}{cccccccccccccc}
\hline
\Tstrut\Bstrut
& $a_\star$ & $\nicefrac{L}{L_\mathrm{Edd}}$ & $\dot{M}$ &$R_\mathrm{cor}$ & $\log(L_\mathrm{AD})$ & $\log(L_\mathrm{SX})$ & $\log(L_\mathrm{PL})$ & $T_\mathrm{tor}$ & $\log(L_\mathrm{tor})$ & $R_\mathrm{tor}$ & $\log(L_\mathrm{AGN})$ & $\chi^2$/dof \\ 
& & & [\nicefrac{M$_{\sun}$}{yr}] & [$R_\mathrm{g}$] &  [erg/s] & [erg/s] & [erg/s] & [K] & [erg/s] & [ld] & [erg/s] &  \\
& (1) & (2) & (3) & (4) & (5) & (6) & (7) & (8) & (9) & (10) & (11) & (12) \vspace{0.1em} \\
\hline
\Tstrut
(a) & 0.0 & 0.93 & 7.3 & 11.7 & 46.64 & 45.53 &  45.17 & 1430 & 46.19 & 4000 & 46.68 & 172/126 \\
(b) & 0.8 & 2.16 & 7.8 & 2.90 & 46.91 & 00.00 &  00.00 & 1350 & 46.21 & 4000 & 46.91 & 198/127 \\

\hline
\end{tabular}
\label{tab:optxconv}
\parbox[]{16.2cm}{\vspace{0.2em} The columns are: (1) dimensionless BH spin: this was fixed at this value; (2) Eddington ratio; (3) mass accretion rate; (4) outer coronal radius in gravitational radii, $R_\mathrm{g}={\color{black}2.95}\times10^{11}$~m $={\color{black}1.14}\times10^{-2}$~light days; (5) luminosity of the accretion disc; (6) luminosity the soft Comptonisation region `SX'; (7) luminosity of the power-law tail `PL'; (8) temperature of the dusty torus; (9) luminosity of the IR radiation from the torus; (10) the dusty torus inner radius in light days; (11) the total AGN luminosity AD$+$SX$+$PL; (12) the $\chi^2$ statistic over the number of degrees of freedom (dof) in the model. $\dot{M}$ and $R_\mathrm{tor}$ are not model parameters but have been derived from our results.}
\end{center}
\end{table*}

\subsubsection{The broad line region luminosity and radius}
The broad line region (BLR) is thought to be a system of gas clouds
distributed on sub-parsec scales, very close to the central BH.
The clouds are photoionised by the intense radiation from the accretion flow
and the resulting line emission (and reflection) from the clouds can be Compton-scattered to higher
energies by interaction with the particles in the relativistic jet.

Following the method described in \cite{Kynoch18}, we estimate the 
luminosity and radius of the BLR, from which the energy density of BLR photons can be calculated.
Our X-shooter spectrum covers four of the
prominent broad emission lines, (\CIII, Mg\,\textsc{ii}, H$\upbeta$ and
H$\upalpha$) from which we can estimate the BLR luminosity. 
The luminosities of the H$\upalpha$ and H$\upbeta$ broad
components are listed in Table \ref{tab:bhmass} and for the Mg\,\textsc{ii} line and
\CIII~broad components we get a luminosity of $\log L_{\rm MgII} =
43.30$~erg~s$^{-1}$ and $\log L_{\rm CIII]} = 43.42$~erg~s$^{-1}$,
  respectively. 
This results in a total BLR luminosity of $\log L_{\rm
    BLR} = 44.54$~erg~s$^{-1}$.
{\color{black} As noted in Section~\ref{sec:bhmass}, we estimate} the BLR radius to be $260^{+91}_{-67}$~light-days 
($\approx{\color{black}2.28}\times10^4~R_\mathrm{g}$ for our adopted BH mass) from the 
H$\upbeta$ radius-luminosity relationship {\color{black} of \cite{Bentz13}}.

\subsubsection{The dusty torus luminosity and radius}
Photons from the extended ($\sim$parsec-scale) dusty torus are also upscattered
by interaction with the jet.
We parameterise this source of seed photons by the torus luminosity and radius 
(the latter is dependent on the dust temperature).
The hot dust luminosity and temperature are determined from the
blackbody fit to the mid-IR photometry: the values are listed in Table \ref{tab:optx}. 
Then, for a silicate dust composition, we estimate the hot
dust radius to be $\sim 4000$ light-days ($\approx{\color{black}3.51}\times10^5~R_\mathrm{g}$).
We note that the torus hot dust radius may be smaller than this depending on the 
assumption of grain size.


\subsection{The broadband SED}
\label{sec:jetmodel}
\begin{figure*}
\centering
	\begin{tabular}{cc}
	\includegraphics[width=\columnwidth, keepaspectratio]{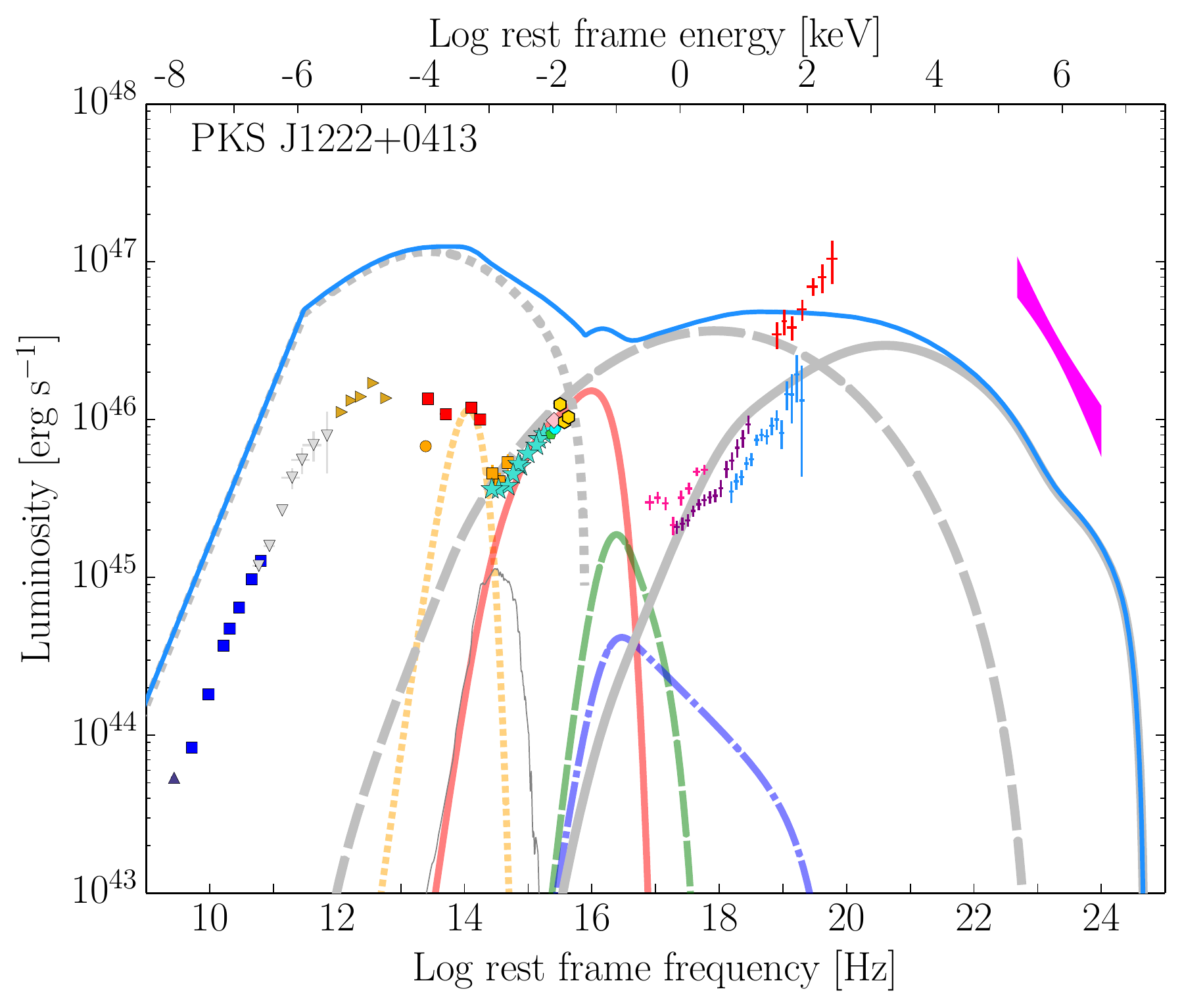} 			&
	\includegraphics[width=\columnwidth, keepaspectratio]{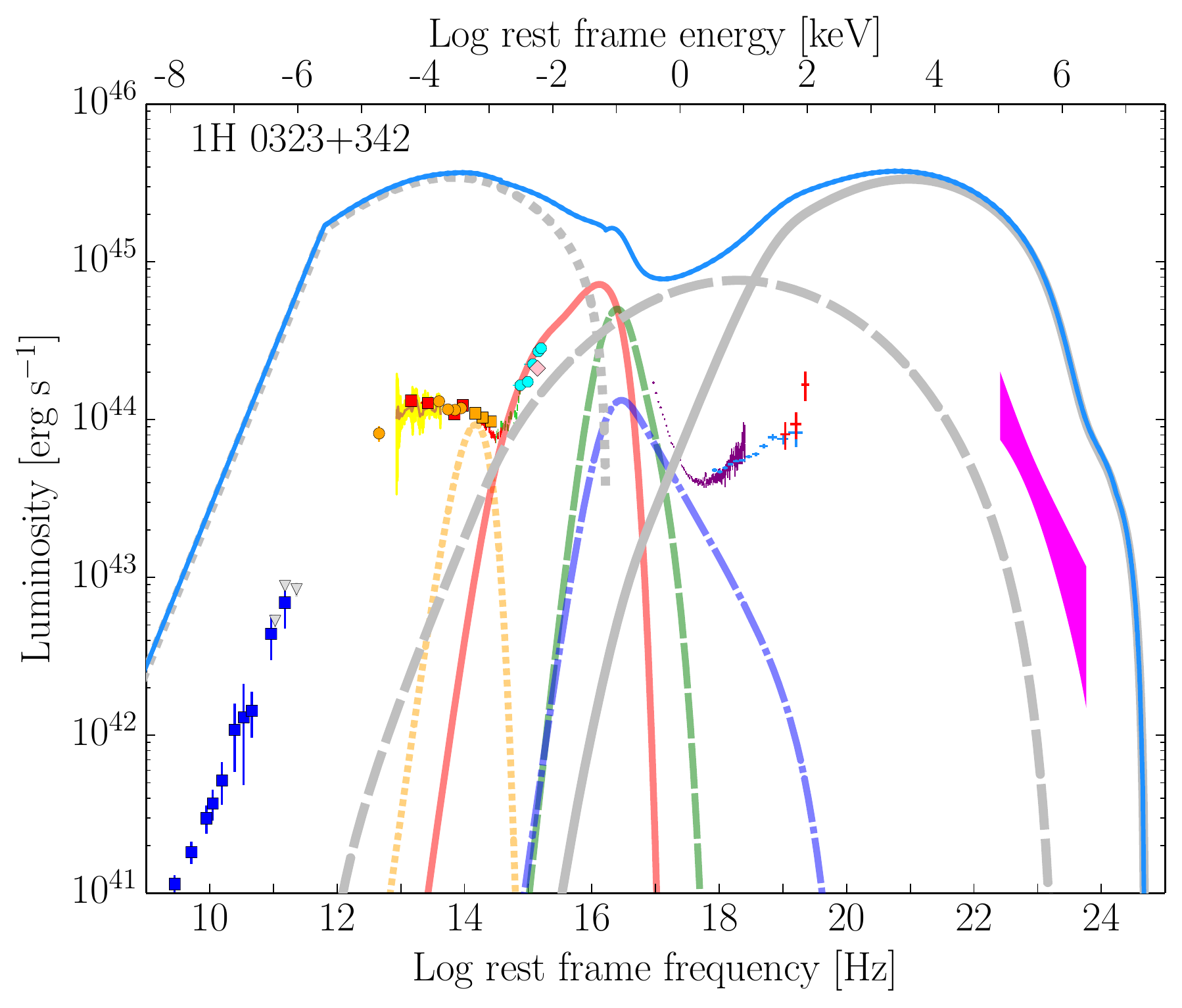} 				\\
	\includegraphics[width=\columnwidth, keepaspectratio]{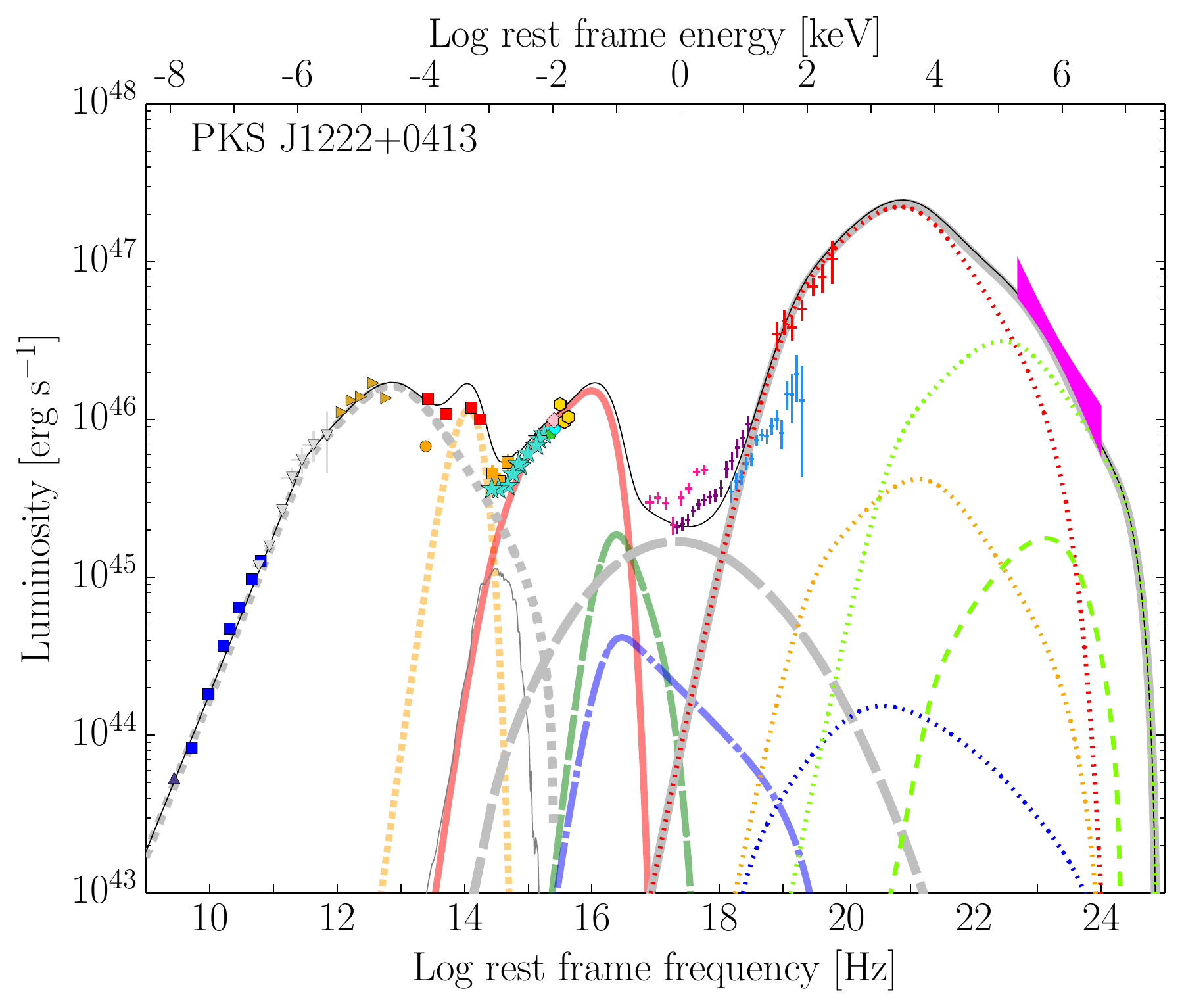} &
	\includegraphics[width=\columnwidth, keepaspectratio]{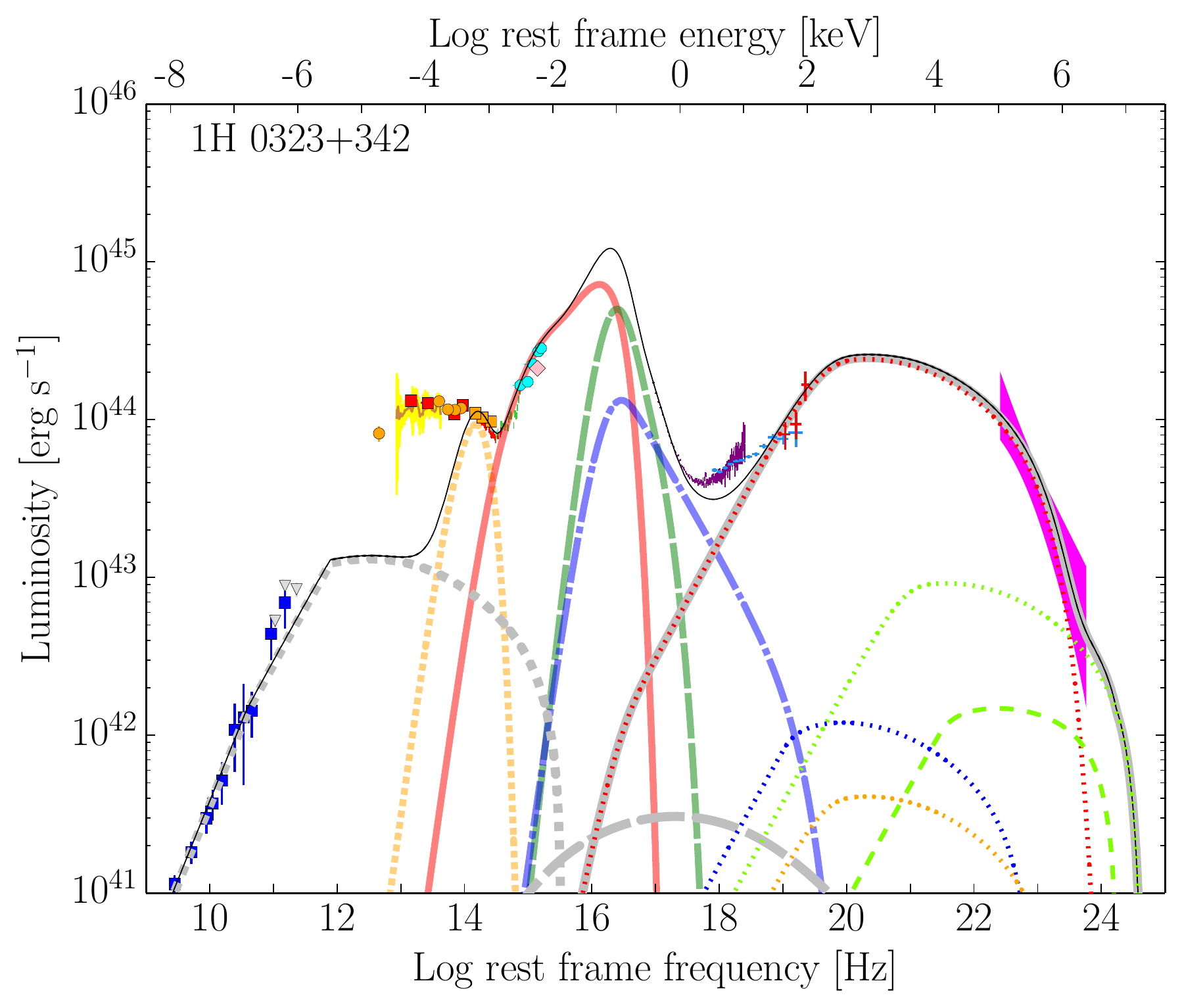} 	\\
	\multicolumn{2}{c}{\includegraphics[width=1.8\columnwidth, keepaspectratio]{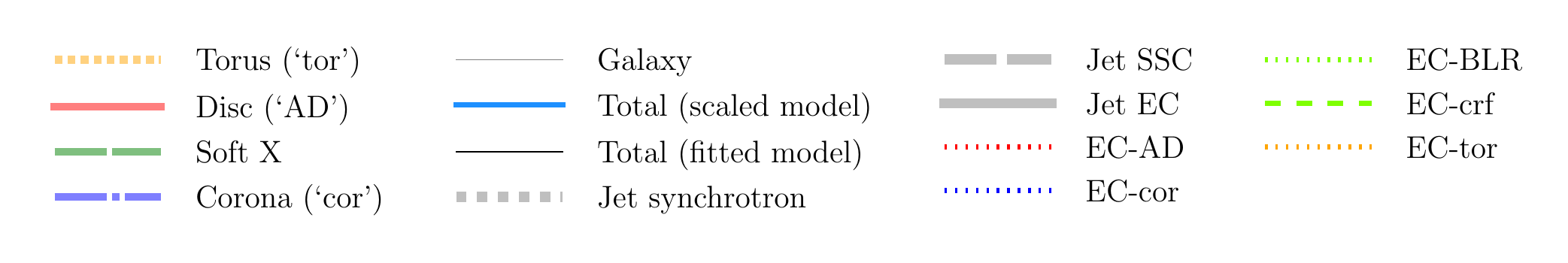}}
	\end{tabular}
\caption[General Caption]{\textit{Top:} Black hole mass and accretion rate scaled models of the jet emission from PKS\,J1222$+$0413 (left) and 1H\,0323$+$342 (right).
\textit{Bottom:} Fitted EC-disc models for PKS\,J1222$+$0413 (left) and 1H\,0323$+$342 (right).  
{\color{black}We show the three components of the jet emission (synchrotron, synchrotron self-Compton `SSC' and external Compton `EC') as grey lines.
The individual EC components (from the disc, corona, BLR, reflection of the corona off the BLR `crf' and torus) are shown as coloured dotted and dashed lines.}
The PKS\,J1222$+$0413 data are coded the same as Figure~\ref{fig:mwl}.
The 1H\,0323$+$342 models were presented by \cite{Kynoch18}.}
\label{fig:jet-scaled}
\end{figure*}

We use the \cite{Gardner18} single-zone, leptonic jet code \textsc{jet} (based on the physics of \citealt{G&T09}), 
adapted so that the external photon field can be set via input parameters\footnote{\color{black}A concise description of the model was given in \cite{Kynoch18}; further details can be found in \cite{Gardner18} and \citealt{G&T09}.}.
We first test the standard jet scalings against our data.
For a BH of mass $M_\mathrm{BH}={\color{black}2}\times10^8$~M$_{\sun}$ with $L/L_\mathrm{Edd}=\dot{m}={\color{black}0.93}$, the scalings of \cite{Gardner18} predict a magnetic field strength of $B\propto\sqrt{\dot{m}/M_\mathrm{BH}}={\color{black}17.7}$~G and power injected into relativistic electrons $P_\mathrm{rel}\propto \dot{m}M_\mathrm{BH}={\color{black}3.72}\times10^{43}$~erg~s$^{-1}$.
Lengths scale linearly with mass, so in mass-normalised units the site of the jet emission region is equal to that of a typical \cite{G&T10} FSRQ, i.e.\ $Z_\mathrm{diss}=1280~R_\mathrm{g}$.  For the jet geometry, we assume an opening angle $\phi=0.1$~radians and inclination to the line of sight equal to the inverse of the bulk Lorentz factor $i=1/\Gamma_\mathrm{BLF}$.
We keep all other jet parameters equal to the mean FSRQ parameters of \cite{G&T10}, listed under the ``Scaled'' model in Table~\ref{tab:jet}, the resultant model is shown in Figure~\ref{fig:jet-scaled}.

We find that the SED is not as discrepant as when performing the same test with 1H\,0323$+$342 (see the top panels of Figure~\ref{fig:jet-scaled}).  
The jet power $P_\mathrm{j}$ (relative to disc power) in this scaled model is comparable to other blazars but clearly the model needs some adjustment to fit the data.

{\color{black}
We adjust the parameters of the scaled model to better fit the broadband SED. 
The Compton dominance (the ratio of peak Compton to synchrotron peak luminosities) of the scaled model is clearly too low; this can be increased by lowering $B$.
The magnetic field is lowered approximately by a factor five in our fitted model compared to the scaled one.  
This is necessary in part to avoid the synchrotron self-Compton peak overpredicting the low-energy X-ray data.
Increasing $P_\mathrm{rel}$ and $\Gamma_\mathrm{BLF}$ has the effect of increasing the jet SED in luminosity.
A solution near equipartition (equal energy densities in electrons and the magnetic field) is found by adjusting $B$, $P_\mathrm{rel}$ and $\Gamma_\mathrm{BLF}$.
By increasing $\Gamma_\mathrm{BLF}$, we can lower $P_\mathrm{rel}$ to reach approximate equipartition, having lowered $B$.
Then, the shapes of the synchrotron and EC peaks are adjusted by tuning the parameters of the injected electron distribution (i.e. the low- and high-energy slopes $s_1$ and $s_2$ and the break Lorentz factor $\gamma_\mathrm{brk}$).  
It is necessary to lower $\gamma_\mathrm{brk}$ slightly to match the frequency peak of the synchrotron emission.
The jet power in this fitted model is approximately half that of $L_\mathrm{AD}$.

It can be seen in Figure~\ref{fig:jet-scaled} that this fitted EC-disc model fits our broadband SED well.  
In particular we are able to achieve good agreement with our optical/UV data from X-shooter and \textit{Swift} UVOT, soft X-ray data from \textit{NuSTAR} and the \textit{Fermi} $\gamma$-ray spectrum, all of which were obtained within the $\sim$three-month period 2017 March--June.
However, this model does not fit the high-energy \textit{NuSTAR} specrtum as well, although it is likely that the spectral shape above $\approx20$~keV is not very well defined by the \textit{NuSTAR} data.  
The harder jet X-ray spectrum in our model does match the \textit{Swift} BAT spectrum taken from the 105-month monitoring campaign.
The model also slightly overpredicts the mid-infrared data.
This may be because we have not accounted for any synchrotron jet emission in the infrared in our modelling in Section~\ref{sec:optxconv}.
The mid-infrared variability seen in the \textit{WISE} lightcurve (Figure~\ref{fig:wiselc}) would suggest that the jet does make some contribution in this waveband.
Any overestimation we may have made of the hot dust luminosity will have minimal impact on the models we have presented.
This is because the jet emission region is closer to the accretion disc and BLR than the torus, therefore the energy density of infrared torus seed photons is relatively low.        
}  

\begin{table*}
\small
\caption{Jet parameters obtained from spectral fits to the full multiwavelength SED with \textsc{bbody}+\textsc{optxconv}+\textsc{jet} models}
\begin{center}
%
\begin{tabular}{lccccc}
\hline
Parameter 										& Units 						& \multicolumn{4}{c}{Model value} 							\\
\hline
\Tstrut\Bstrut									&								& \multicolumn{2}{c}{PKS\,J1222$+$0413} & \multicolumn{2}{c}{1H\,0323$+$342} 	\\
												&								& \multicolumn{2}{c}{$\overbrace{\rule{3.5cm}{0pt}}$}	& \multicolumn{2}{c}{$\overbrace{\rule{3.5cm}{0pt}}$}       \\ 		
												&								& Scaled with				& EC-disc   & Scaled with	& EC-disc							\\
												&								& $M_\mathrm{BH}$ \& $\dot{m}$	&   & $M_\mathrm{BH}$ \& $\dot{m}$	& 							\\
												
\multicolumn{6}{l}{Input parameters:}																								        \\																							
$Z_\mathrm{diss}$								& [$R_\mathrm{g}$] (ld)			& 1280 ({\color{black}15})	& 1280 ({\color{black}15})	& 1280 (1.5) 	& 1280 (1.5) \\
$a_\star$										&								& {\color{black}0.0}			& {\color{black}0.0}			& 0.0  			& 0.0				 \\
$i$												& [deg]							& 4.41 	 					& {\color{black}2.86}			& 4.41 			& 4.77				  \\
$\Gamma_\mathrm{BLF}$							&								& 13						& {\color{black}20}			& 13 			& 12				  \\
$\delta$										&								& 13	 					& {\color{black}20}			& 13			& 12				 \\
$B$												& [G]							& {\color{black}17.7}			& {\color{black}3.75}			& 38			& 8				   	\\
$\log(P_\mathrm{rel})$							& [erg\,s$^{-1}$]				& {\color{black}43.6}			& {\color{black}42.9}			& 42.24			& 41.0				 \\
$\gamma_\mathrm{min}$							&								& 1.00						& 1.00						& 1.00			& 1.00				  \\
$\gamma_\mathrm{brk}$							&								& 300	 					& 200						& 300			& 300				   \\
$\gamma_\mathrm{max}$							& 								& 3000						& 3000						& 3000			& 3000				 \\
$s_1$											&								& 1.0	 					& -1.0						& 1.0			& 1.5				  \\
$s_2$											&								& 2.7	 					& {\color{black}3.2}			& 2.7			& 2.7				   \\
\multicolumn{6}{l}{Output / derived parameters:}																								        \\	
$\gamma_\mathrm{cool}$							&								& {\color{black}12}			& {\color{black}23}			& 19			& 47				   \\
$\log(\nu_\mathrm{ssa})$						& [Hz]							& {\color{black}11.5}			& {\color{black}11.5}			& 11.6			& 10.6				 \\
$\log\left(\nu_\mathrm{peak}^\mathrm{sync}\right)$			& [Hz]				& {\color{black}13.5}			& {\color{black}12.9}			& 13.8			& 12.5				  \\
$\log\left(\nu L_{\nu_\mathrm{peak}}^\mathrm{sync}\right)$	& [erg\,s$^{-1}$]	& {\color{black}47.1}			& {\color{black}46.2}			& 45.56			& 43.12				   \\

$\log(P_\mathrm{rad})$							& [erg\,s$^{-1}$]				& {\color{black}45.32}		& {\color{black}44.94}		& 43.95 		& 42.51  			\\
$\log(P_\mathrm{e})$							& [erg\,s$^{-1}$]				& {\color{black}45.01}		& {\color{black}44.69}		& 43.76			& 42.74				  \\
$\log(P_B)$										& [erg\,s$^{-1}$]				& {\color{black}45.45}		& {\color{black}44.48}		& 44.12			& 42.70				  \\
$\log(P_\mathrm{p})$							& [erg\,s$^{-1}$]				& {\color{black}47.28}		& {\color{black}46.31}		& 45.91			& 45.01				  \\
$\log(P_\mathrm{j})$							& [erg\,s$^{-1}$]				& {\color{black}47.30}		& {\color{black}46.35}		& 45.93			& 45.01				   \\
$P_\mathrm{j}/L_\mathrm{AD}$					& 								& {\color{black}4.4}			& {\color{black}0.51}			& 4.3			& 0.52				 	 \\
$U_\mathrm{e}/U_B$								&								& {\color{black}0.37}			& {\color{black}1.6}			& 0.44			& 1.1				   	\\

\hline
\end{tabular}
\label{tab:jet}
\parbox[]{11cm}{\vspace{0.2em} Comparison of jet models for PKS\,J1222$+$0413 and the best fit model for 1H\,0323$+$342 determined by \cite{Kynoch18}.  $P_\mathrm{rad}$, $P_\mathrm{e}$, $P_B$, $P_\mathrm{p}$ and $P_\mathrm{j}$ are the radiative, electron, magnetic field, kinetic and total jet powers, respectively.
$U_\mathrm{e}/U_B$ is the ratio of relativistic electron to magnetic field energy densities.
The corresponding models are shown in Figure~\ref{fig:jet-scaled}.}
\end{center}
\end{table*}

\section{Discussion}
\subsection{PKS\,J1222$+$0413 as a NLS1}
\label{sec:nls1}
In Section~\ref{sec:bhmass} we found that the FWHM of the H$\upalpha$ and H$\upbeta$ Balmer lines and the {\color{black}individual lines of the} Mg\,\textsc{ii} {\color{black}doublet are each lower than} 2000~km~s$^{-1}$, the limiting line width that is usually used to separate NLS1s from broad-line Seyferts (see Table~\ref{tab:bhmass}).
{\color{black} 
If, however, we use Gaussian rather than Lorentzian profiles to model the Mg\,\textsc{ii} lines, we determine FWHMs of 2240~km~s$^{-1}$.
\cite{Yao15} found a similar $\mathrm{FWHM}=2264$~km~s$^{-1}$ for H$\upbeta$ when modelling its broad component with a Gaussian.
\cite{Torrealba12} modelled the total Mg\,\textsc{ii} profile with two Gaussians (one broad and one narrow) centred at 2800~\AA.  
For the broad component they determined a $\mathrm{FWHM}=5268$~km~s$^{-1}$.
Clearly, the determination of the emission line FWHMs (hence the classification of this source as either a narrow- or broad-line Seyfert 1) depends on the assumed line shape and the contribution of the narrow line component to the total line flux.
Here, we have measured the Balmer line widths directly from their observed profiles, without assuming a particular line shape.
In doing so, we have assumed that the flux contribution of the narrow line region is negligible, based on our estimation from a typical ratio of [O\,\textsc{iii}]$\lambda5007$ to narrow H$\upbeta$ luminosities.
If the narrow Balmer lines are somewhat stronger than we have estimated (which is possible, given the range of values of 
$L_{\mathrm{[O\textsc{iii}]}}/L^{\mathrm{n}}_{\mathrm{H}\upbeta}$ 
determined from large AGN samples\footnote{{\color{black}Note that \cite{Zhou06} found that only $\sim15$ per cent of NLS1s had a ratio $\lesssim5$, and in most of the sources with low ratios a large fraction of the emission was from H\,\textsc{ii} regions of the host galaxies.  It is therefore very unlikely that $L^{\mathrm{n}}_{\mathrm{H}\upbeta}$ is much greater than we have estimated.}}) then the FWHMs of the broad Balmer would correspondingly increase to slightly above 2000~km~s$^{-1}$. 
Alternative measures of the emission line FWHMs of this source may give values exceeding 2000~km~s$^{-1}$, in which case PKS\,J1222$+$0413 is not strictly a NLS1.  
}
{\color{black}However, we point out that the 2000~km~s$^{-1}$} limit is arbitrary and other authors have used more relaxed definitions of NLS1s {\color{black}(e.g.\ \citealt{Zhou06}, \citealt{Rakshit17}, \citealt{Lakicevic18})}.

\cite{Czerny18} emphasise that the 2000~km~s$^{-1}$ division is a \textit{phenomenological} one. 
The NLS1 definition of \cite{OP85} was determined from low-mass objects in the local Universe.
It has been proposed that NLS1s are objects that lie at the extreme of `Eigenvector 1' (\citealt{Bor92}), with the sequence driven primarily by the accretion rate.
A high accretion rate may be a better indicator of the high-mass, high-$z$ analogues of local NLS1s than the emission line FWHMs given that linewidths will be greater for higher-mass sources. 
\cite{Czerny18} calculate that the dividing FWHM between {\color{black}narrow- and broad-line AGN} is a factor $\approx2$ greater in objects of a mass $3\times10^8$~M$_{\sun}$ than those of $10^7$~M$_{\sun}$.
Therefore for PKS\,J1222$+$0413, based on the BH mass the above scaling would imply a FWHM $\lesssim4000$~km~s$^{-1}$.
Additionally, the strong Fe\,\textsc{ii} and weak [O\,\textsc{iii}] emission seen in the optical spectra are typical of NLS1s.
We therefore confirm that PKS\,J1222$+$0413 is a NLS1, although one with a much greater than average mass.

The apparent weakness of the NLR is consistent with the view of this object as a high-luminosity, highly-accreting quasar.  
\cite{Netzer04} suggested that the NLRs of high-luminosity AGN may be different to those of lower-luminosity, nearby sources.
The NLR of high-luminosity sources may be lost as a result of being dynamically unbound.
\cite{Collinson17} found an anti-correlation between [O\,\textsc{iii}]\,$\lambda5007$ line strength and Eddington ratio in their
sample of high-redshift AGN.
Several of their highest accretion rate sources had very weak [O\,\textsc{iii}] emission, similar to what we see in PKS\,J1222$+$0413.
Based on their study of the super-Eddington NLS1 RX\,J0439.6$-$5311, \cite{Jin17} present a picture of high accretion rate AGN which explains the weakness of the NLR as the result of it being (partly) shielded from the nuclear ionising flux by a `puffed up' inner accretion disc or by a disc wind. 
Given the Eddington ratio determined for this source, it is unlikely that PKS\,J1222$+$0413 has a super-Eddington wind like RX\,J0439.6$-$5311.
However, its SED peaks in the FUV and the accretion flow X-ray spectrum is very soft {\color{black} and would therefore not over-ionise material driven out in a UV line-driven wind.}  
It may be that PKS\,J1222$+$0413 has a UV line-driven wind partly shielding the NLR.
Although the narrow emission lines are weak, we find that the high-ionisation lines (emitted by gas between the BLR and NLR) are strong, e.g.\ Neon {\color{black}([Ne\,\textsc{iv}]$\lambda2423$)} and the Iron transitions {\color{black}(Fe\,\textsc{xi})}.

\subsection{The jet of PKS\,J1222$+$0413}
It was challenging to extract the $\gamma$-ray spectrum of PKS\,J1222$+$0413 (Section~\ref{sec:fermi}).
Because of the close proximity of the bright $\gamma$-ray sources 3C\,273 and PKS\,1237$+$049 we found that very few photons in the field had a strong statistical association with PKS\,J1222$+$0413; most were more favourably associated with the other sources.
Applying an 85~per~cent confidence cut, we are left with very few photons and a limited $\gamma$-ray spectrum.
As we noted earlier, there have been no $\gamma$-ray flares recorded by \textit{Fermi} near the location of PKS\,J1222$+$0413. 
Rapid variability in the optical has not been recorded either, with \cite{Ojha18} noting that the intra-night optical variability was remarkably low, and consistent with levels see in non-jetted (radio-quiet) sources.
However, the \textit{WISE} data (Figure~\ref{fig:wiselc}) show variability which is likely from jet synchrotron emission.

Whereas the viewing angle to the jet (and hence the bulk Lorentz and Doppler factors, under the assumption $i=1/\Gamma_\mathrm{BLF}\approx\delta$) was constrained by radio measurements (\citealt{Fuhrmann16}) in the case of 1H\,0323$+$342 (\citealt{Kynoch18}), we have no such observational constraint for the jet of PKS\,J1222$+$0413. 
\cite{Lister16}\footnote{In this paper our source is named 4C\,$+$04.42.} investigated the radio jet of PKS\,J1222$+$0413 as part of the MOJAVE project and found that the jet is two-sided on kpc scales (but it is one-sided on pc scales).
\cite{Lister18} notes that no measurable motion within the jet has been recorded in a 14-year monitoring period and that the maximum apparent jet speed is sub-luminal: $(0.9\pm0.3)c$.  
If the transverse speed of the jet $v_\mathrm{t}=0.9c$, this would imply a large viewing angle to the jet axis with $i\approx50$--90$^\circ$ and a low bulk Lorentz factor whereas high Lorentz factors are determined by SED modelling (e.g.\ \citealt{Yao15}).
The viewing angle is likely to be small because of the one-sided pc-scale jet structure.
Its long-term (2008 to date) 15~GHz radio lightcurve, recorded by the Owens Valley Radio Observatory (OVRO, \citealt{Richards11}), shows variability which is also suggestive of a high Doppler factor and low viewing angle.  
It is likely that the apparent sub-luminal motion is due to a lack of traceable bright knots, rather than a low-speed jet.

Whilst it is difficult to extract a reliable $\gamma$-ray spectrum for PKS\,J1222$+$0413 because of its low $\gamma$-ray flux and the presence of bright nearby sources, we are confident of the $\gamma$-ray nature of our source.
Other, well-defined properties such as the radio and infrared variability and high-resolution radio images of the jet are supportive of the presence of very high energy processes. 
Furthermore, the X-ray data from \textit{NuSTAR} and \textit{Swift} BAT are evidence of a high-energy Compton component, and the $\gamma$-ray spectrum appears to be consistent with the very high-energy tail of this feature. 

\subsection{Comparison with 1H\,0323$+$342}
We presented a very similar analysis of the lowest-redshift $\gamma$-NLS1, 1H\,0323$+$342, in \cite{Kynoch18}.
Here we make a comparison with our findings from PKS\,J1222$+$0413.  
Obviously, PKS\,J1222$+$0413 {\color{black} has a much greater BH mass} than 1H\,0323$+$342.
We therefore expect PKS\,J1222$+$0413 to be more similar to a standard (high-mass) blazar.
Indeed, when we scale a standard FSRQ jet by mass and mass accretion rate we see that discrepancy between the predicted SED and the data is much less than \cite{Kynoch18} found for 1H\,0323$+$342.
The high accretion rate of PKS\,J1222$+$0413 means that a high magnetic field is predicted from the scaling relations.
As with 1H\,0323$+$342, the resultant predicted SED has a low Compton dominance and a shape much more akin to a BL Lac, albeit at a much higher luminosity than BL Lacs.
We can better fit the data by turning down the magnetic field strength.
(With 1H\,0323$+$342 it was necessary to turn down both the magnetic field strength and the power injected into the relativistic electrons.)
As with 1H\,0323$+$342, a typical NLS1 SED model fits the observed NIR--X-ray SED well, if a (jet) power-law is added to account for hard X-rays.
In 1H\,0323$+$342 \cite{Kynoch18} found that the X-ray photon index below $\approx2$~keV was very soft, typical of (non-jetted) NLS1s, whereas above 2~keV, the photon index was much harder, typical of nonthermal jet emission.
This was interpreted as a transition from accretion-flow-dominated to jet-dominated emission.
We do not find {\color{black} such strong} evidence of a soft, low-energy photon index in PKS\,J1222$+$0413 (Section{\color{black}s}~\ref{sec:xmm} {\color{black}and \ref{sec:optxconv}}), likely because its higher mass and redshift mean that the accretion flow makes less of a contribution to the flux in the \textit{XMM-Newton} bandpass.

\cite{Kynoch18} showed that 1H\,0323$+$342 had a low jet power for its accretion power, compared to the \textit{Fermi} blazars studied by \cite{Ghisellini14}.
Here, we find that {\color{black} the jet of} PKS\,J1222$+$0413 {\color{black} is similarly underpowered relative} to its accretion power.  {\color{black}However, since it has} a BH mass similar to FSRQs, {\color{black} in terms of its {\it absolute} jet power} it occupies the same region of the {\color{black} blazar sequence} as the FSRQs, as one might expect. 
\cite{Foschini17} proposed that there are two branches on the blazar sequence.
This can be seen if the jet power (calculated from the radio luminosity) is plotted against the disc luminosity. 
The track of the main branch, containing FSRQs and BL Lacs (which have similar BH masses), may be explained as a difference in the environment in which the relativistic electrons cool.
The lower track, connecting radio-loud NLS1s to FSRQs (which have similar cooling environments), must be driven by the difference in the BH masses.
{\color{black}A small number of sources have jet powers estimated both from their radio luminosity and from SED modelling.
We find that the jet power calculated via SED modelling is generally greater than that estimated from the radio luminosity.} 
\cite{Paliya19} have modelled the broadband SEDs of sixteen $\gamma$-NLS1s to determine the jet powers, and have compared these to a sample of previously-modelled blazars.
Their results suggest that $\gamma$-NLS1s form the low-jet-power tail of the FSRQ population, rather than forming a separate branch.

{\color{black} \cite{Yao15} also modelled the SED of PKS\,J1222$+$0413 and presented both an EC-BLR and EC-torus case in which
the $\gamma$-rays were predominantly upscattered seed photons from the BLR or torus, respectively.
Here, we have demonstrated that an EC-disc model can also reproduce the observed SED.
In our EC-disc model we determine a high bulk Lorentz factor $\Gamma_\mathrm{BLF}=20$, which is substantially greater than the mean value 13 \cite{G&T10} found for FSRQs and the value 12 \cite{Kynoch18} found for 1H\,0323$+$342. 
The factors determined by \cite{Yao15} from their jet models were even higher: $\Gamma_\mathrm{BLF}=26$, 35 for the EC-BLR and EC-torus models, respectively.
Unfortunately, \cite{Yao15} do not quote any jet powers determined from their models, or give sufficient details of their model parameters and so it is not possible to undertake a meaningful comparison. 
We emphasise that}
the jet power and parameters determined by SED fitting are still model-dependent.
The results which are obtained will depend on the detailed assumptions made, and the approaches taken to compute the complex physical processes. 
Even within the same prescription, the complexity of jet models mean that it is difficult to converge on a unique solution.
Our approach has been to construct a quasi-simultaneous data set and to self-consistently model and apply the ambient photon field which is crucial in reproducing the observed external Compton (X-ray and $\gamma$-ray) emission.
In future studies it would be useful if the results were presented in a more uniform way, with equivalent model parameters and basic assumptions stated.  


\section{Conclusions}
Our main results are summarised below:
\begin{itemize}
\item The BH mass ${\color{black}\approx2}\times10^{8}$~M$_{\sun}$ is very high for a NLS1, which typically have an order of magnitude smaller BH masses.
\item {\color{black}We conclude that PKS\,J1222$+$0413 is a NLS1, based on our new data and analysis of its optical spectral properties and the accretion flow parameters obtained from modelling its broadband SED.
This is in agreement with the NLS1 classification made by \cite{Yao15}.}
\item The very weak [O~\textsc{iii}] is suggestive of a UV line-driven disc wind shielding the narrow-line region from ionising UV flux produced near to the BH. 
\item The jet power is {\color{black} approximately half of} the accretion disc luminosity {\color{black} whereas for BL Lac and FSRQ-type blazars, the jet power typically exceeds the accretion power ({\citealt{Ghisellini14})}.
This finding is similar to the result for the $\gamma$-NLS1} 1H\,0323$+$342 which \cite{Kynoch18} also found to have an underpowered jet. 
\item Because of its high BH mass, similar to FSRQ-type blazars, PKS\,J1222$+$0413 lies in the region of the {\color{black} blazar sequence} occupied by the FSRQs.  
Most other $\gamma$-ray detected NLS1s, like 1H\,0323$+$342, have much lower {\color{black}absolute} jet powers (similar to BL Lacs) but share many of the same jet properties as FSRQs.
{\color{black}PKS\,J1222$+$0413 is therefore an ideal object in which to explore the relationship between $\gamma$-NLS1s and FSRQs.}    
\end{itemize}

\section*{Acknowledgements}
DK acknowledges the receipt of an STFC studentship (ST/N50404X/1). 
DK, HL, MJW and CD acknowledge support from the STFC (ST/P000541/1).
Thanks to Gerard Kriss for providing electronic versions of the \cite{Telfer02} composite spectra.
Thanks to Luigi Foschini and Matt Lister for useful discussions.
In this paper we have made use of the following:
\begin{itemize}
\item the ESO \textsc{reflex} data reduction software (\citealt{Reflex});
\item data from \textit{Fermi}, a NASA mission operated and funded NASA, the U.S. Department of Energy and institutions in France, Germany, Japan, Italy and Sweden;
\item data from \textit{Spitzer}, \textit{WISE}, \textit{Planck} and 2MASS which are available from the NASA / IPAC Infrared Science Archive, which is operated by the Jet Propulsion Laboratory, California Institute of Technology, under contract with NASA;
\item data from \textit{NuSTAR}, a project led by the California Institute of Technology, managed by the Jet Propulsion Laboratory and funded by NASA;
\item data from \textit{Swift}, and its XRT Data Analysis Software (XRTDAS) developed under the responsibility of the ASI Science Data Center (ASDC), Italy;
\item data from and software developed for \textit{XMM-Newton}, an ESA science mission with instruments and contributions directly funded by ESA Member States and NASA; 
\item data and software (including the ftools\footnote{\url{http://heasarc.gsfc.nasa.gov/ftools/}} \citealt{Blackburn95}) provided by the High Energy Astrophysics Science Archive Research Center (HEASARC), which is a service of the Astrophysics Science Division at NASA/GSFC and the High Energy Astrophysics Division of the Smithsonian Astrophysical Observatory;
\item data from SDSS: funding for the SDSS, SDSS-II and SDSS-III has been provided by the Alfred P.\ Sloan Foundation, the Participating Institutions, the National Science Foundation, the U.S.\ Department of Energy, NASA, the Japanese Monbukagakusho, the Max Planck Society, and the Higher Education Funding Council for England;
\item Ned Wright's cosmology calculator (\citealt{Wright06});
\item Doug Welch's excellent absorption law calculator (\url{http://www.dougwelch.org/Acurve.html}).
\end{itemize}




\bibliographystyle{mnras}
\bibliography{bib-pksj1222} 






\bsp	
\label{lastpage}
\end{document}